\documentclass[11pt, a4paper]{article}
\pdfoutput=1
\usepackage{graphicx}
\usepackage{amssymb}
\usepackage{amsmath}
\usepackage{bm}
\usepackage[table,dvipsnames]{xcolor} 
\usepackage{cite}
\usepackage{slashed}
\usepackage{epstopdf}            
\usepackage{epsfig}
\usepackage{here}
\usepackage{comment}
\usepackage{booktabs} 
\usepackage{colortbl} 
\usepackage{wrapfig}
\usepackage{ascmac}
\usepackage{fancybox}  
\usepackage{subcaption}
\usepackage{soul} 
\usepackage{tikz}
\usepackage{tikz-feynman}
\tikzfeynmanset{compat=1.1.0}
\usetikzlibrary{positioning} 
\renewcommand{\thefootnote}{\fnsymbol{footnote}}
\numberwithin{equation}{section} 
\def\beq#1\eeq{\begin{align}#1\end{align}}
\newcommand{\ov}{\overline}

\newcommand{\1}{\mbox{1}\hspace{-0.25em}\mbox{l}}

\def\vev#1{\langle #1\rangle}
\def\VEV#1{\left\langle #1\right\rangle}

\newcommand{\hc}{{\rm h.c.}}
\newcommand{\eff}{{\rm eff}}

\newcommand{\DM}{{\rm DM}}

\newcommand{\GeV}{{\rm GeV}}
\newcommand{\TeV}{{\rm TeV}}
\newcommand{\MeV}{{\rm MeV}}

\newcommand{\eV}{{\rm eV}}

\newcommand{\lam}{\lambda}

\newcommand{\ol}[1]{\overline{#1}}
\newcommand{\wt}[1]{\widetilde{#1}}

\renewcommand{\[}{\left[}
\renewcommand{\]}{\right]}
\renewcommand{\(}{\left(}
\renewcommand{\)}{\right)}

\RequirePackage{xspace}
\def\Bbar    {\kern 0.18em\overline{\kern -0.18em B}{}\xspace}

\usepackage{caption}
\definecolor{BlueViolet}{rgb}{0.2, 0.00, 0.7}
\definecolor{Blue}{rgb}{0.15, 0.00, 0.9}
\definecolor{light_blue}{rgb}{0.15, 0.35, 0.95}
\definecolor{kit_green}{rgb}{0, 
0.58823 
, 0.50980 
}
\usepackage[
colorlinks=true, linkcolor=light_blue,citecolor=light_blue,urlcolor=kit_green]{hyperref}

\graphicspath{{figs/}} 	     

\begin{document}
\sloppy 

\begin{titlepage}

\begin{center}

\hfill{YITP--25--160}\\
\vskip .3in

{\Large{\bf Dirac neutrino and dark matter \\ in left-right symmetric models}} \\

\vskip .3in

\makeatletter\g@addto@macro\bfseries{\boldmath}\makeatother

{\large 
Shohei Okawa$^{\rm a,b,c}$, Yuji Omura$^{\rm d}$ and Keyun Wu$^{\rm e}$}
\vskip .25in
$^{\rm (a)}${\it Asia Pacific Center for Theoretical Physics, Pohang, 37673, Korea}\\
$^{\rm (b)}${\it Department of Physics, Pohang University of Science and Technology, Pohang, 37673, Korea}\\
$^{\rm (c)}${\it Yukawa Institute for Theoretical Physics, Kyoto University, Kyoto 606-8502, Japan}\\
$^{\rm (d)}${\it Department of Physics, Kindai University, Higashi-Osaka, Osaka 577-8502, Japan} \\
$^{\rm (e)}${\it
Departament de F\'isica Qu\`antica i Astrof\'isica, Institut de Ci\`encies del Cosmos (ICCUB),
Universitat de Barcelona, Mart\'i i Franqu\`es 1, E-08028 Barcelona, Spain
}\\[3pt]

\end{center}
\vskip .3in

\begin{abstract}
We study neutrino mass generation and dark matter in a left-right symmetric model. 
The model is based on an $SU(3)_c\times SU(2)_L \times SU(2)_R \times U(1)_{B-L}$ gauge theory with a softly broken parity symmetry. 
Masses of the charged leptons and neutrinos are generated radiatively at one-loop and three-loop level respectively, through their interactions with newly introduced neutral fermion and scalar particles. 
A mass hierarchy of those new particles is required to reproduce the observed patterns of the charged lepton spectrum and neutrino oscillation data. 
The resulting light particles, whose mass can be as light as GeV, serve as good dark matter candidates. 
The phenomenology of such dark matter candidates is governed by their interactions to left- or right-handed neutrinos. 
We study physics of dark matter with several benchmark parameter sets that reproduce the realistic neutrino mass matrix structure, and identify viable parameter spaces. 
\end{abstract}
\vspace{0.3cm}
\end{titlepage}

\setcounter{page}{1}
\renewcommand{\thefootnote}{\#\arabic{footnote}}
\setcounter{footnote}{0}

\hrule
\tableofcontents
\vskip .2in
\hrule
\vskip .4in


\section{Introduction}
\label{sec:intro}

Non-observation of $CP$ violation in quantum chromodynamics (QCD) is a puzzling question in the Standard Model (SM). 
The SM gauge symmetries allow the term, ${\cal L}_\theta=\frac{\alpha_s}{8\pi} \theta G_{\mu\nu}\wt{G}^{\mu\nu}$, where $G_{\mu\nu}$ denotes the gluon field strength tensor and $\wt{G}^{\mu\nu}$ its dual. 
This term violates the parity ($P$) and time-reversal ($T$) symmetries, the latter of which means the violation of the $CP$ symmetry through the $CPT$ theorem. 
The non-vanishing $\theta$ value leads to a nonzero neutron electric dipole moment (EDM), which has not been confirmed by experiments up to now, providing a strong upper limit $|\theta|\lesssim 10^{-10}$. 
The SM has no built-in mechanism to suppress the $\theta$ parameter, and this unexplained smallness is known as the strong $CP$ problem.

Left-right (LR) symmetric theories are a compelling solution to the strong $CP$ problem \cite{Beg:1978mt, Mohapatra:1978fy, Babu:1989rb, Barr:1991qx} (see also Refs.~\cite{Chakdar:2013tca, DAgnolo:2015uqq, Hall:2018let, Craig:2020bnv} for the recent exploration).
Conventionally, LR symmetric models build on an extended gauge group, 
$G_{LR} = SU(3)_c \times SU(2)_L \times SU(2)_R \times U(1)_{B-L}$ \cite{Pati:1974yy,Mohapatra:1974hk,Mohapatra:1974gc,Senjanovic:1975rk,Mohapatra:1980yp}, where $B$ and $L$ denote the baryon and lepton number respectively. 
All SM fermions are embedded in $SU(2)_L$ and $SU(2)_R$ doublet spinor fields, accompanied with right-handed neutrinos. 

In the LR symmetric models, one tends to be faced with a difficulty in reproducing the observed fermion masses and mixing. 
If the fermions obtain their masses and mixings via the couplings to a bidoublet Higgs field \cite{Mohapatra:1980yp}, large flavor changing neutral current (FCNC) processes mediated by extra Higgs bosons bring strong constraints to us, see e.g. Refs. \cite{Iguro:2021nhf,Iguro:2018oou}. 
Moreover, the Higgs potential generally has CP phases in this realization, giving rise to a non-vanishing phase in the vacuum unless there is a symmetry or mechanism that forbids such CP phases \cite{Kuchimanchi:1995rp, Mohapatra:1995xd, Mohapatra:1996vg}, which potentially revives the strong CP problem. 

To avoid these problems, one can instead introduce one $SU(2)_L$ doublet scalar $H_L$ and one $SU(2)_R$ doublet scalar $H_R$, which break the electroweak (EW) symmetry and $SU(2)_R \times U(1)_{B-L}$ respectively. 
The observed quark and charged lepton masses are generated through their couplings to extra $SU(2)_L \times SU(2)_R$ singlet vectorlike quarks and leptons \cite{Balakrishna:1988bn,Babu:1988yq}.
Neutrinos are Dirac fermions in the minimal setup, and their masses are generated at two-loop level.\footnote{Different setups for fermion masses are discussed, for instance, in Refs. \cite{Ma:2017kgb,Ma:1989ap,Bonilla:2023wok,Hall:2023vjb,Gabrielli:2016vbb,deAlmeida:2010qb,Brdar:2018sbk,C:2024exl,Nomura:2016run}. In Ref. \cite{Babu:2024glr}, leptogenesis in the LR symmetric model with Dirac neutrinos is also studied.}
In Refs. \cite{Bolton:2019bou,Babu:2022ikf}, 
the authors analytically perform the two-loop calculation for the neutrino mass generation, and 
show that the model can reproduce the observed neutrino oscillation data in some parameter regions. 

In light of the successful fermion mass generation and the capability of solving the strong CP problem, 
it may be a natural question whether these LR symmetric models can also address the dark matter (DM) problem, one of the biggest empirical puzzles in the SM.\footnote{LR symmetric extensions addressing the DM problem are discussed in Refs.~\cite{Borah:2017leo,Biswas:2024wbz}.}

In this paper, we modify the lepton sector of the model studied in Refs. \cite{Bolton:2019bou,Babu:2022ikf} in order to accommodate the fermion mass generation and DM candidates in a coherent framework. 
Our model features vectorlike neutral fermions $n^a_{L,R}$, and extra $SU(2)_L$ doublet scalars $\wt{L}_L^\alpha$, and $SU(2)_R$ doublet scalars $\wt{L}_R^\alpha$, instead of the vectorlike charged leptons, which forms the main difference from the previous works \cite{Bolton:2019bou,Babu:2022ikf}. 
Couplings of these new fields to the $SU(2)_L$ and $SU(2)_R$ doublet leptons generate the lepton and neutrino masses at one-loop and three-loop levels, respectively. 
Interestingly, in this setup the neutrino mass matrix tends to align with the charged lepton mass matrix. 
Such alignment precisely happens when the neutral scalars originating from $\wt{L}_{L,R}^\alpha$ exhibit a mass spectrum and mixing pattern similar to their charged partners. 
Thus, we need to accommodate different structures with the neutral and charged scalars. 
In this work, we require a mass hierarchy only in the neutral scalar sector to break the alignment, allowing us to reproduce both observed charged lepton masses and neutrino oscillation data. 
As a result of this implementation, a couple of light neutral particles appear in the model. 
The lightest of these light species is stable, thanks to an accidental global symmetry, and therefore serves as a good DM candidate.

This paper is organized as follows.
We first introduce our model and outline how to generate the observed masses and mixings for the quarks and leptons in Sec.\,\ref{sec:model}.
To maintain generality of discussion, we do not fix the number of flavors of newly introduced fields. 
In Sec.\,\ref{sec:illustration}, we demonstrate in a minimal setup that our model can fit the experimental data and predict a neutral stable particle, which plays a role of DM.
After giving a brief comment on cosmological bounds in our model in Sec.\,\ref{sec:Neff}, 
experimental constraints and properties of the DM candidates are discussed in Sec.\,\ref{sec:DM}. 
We summarize our findings in Sec.\,\ref{sec:Summary}.
The detail of the neutrino mass calculation is provided in Appendix \ref{appendix}.

\section{Fermion mass generation in left-right symmetric models}
\label{sec:model}

We consider a LR symmetric extension of the SM. 
Following the conventional setup \cite{Pati:1974yy, Mohapatra:1974hk, Mohapatra:1974gc, Senjanovic:1975rk, Mohapatra:1980yp}, 
the hypercharge gauge group $U(1)_Y$ is extended to $SU(2)_R\times U(1)_{B-L}$. 
The $SU(2)_L$ singlet quark and lepton fields in the SM are embedded into $SU(2)_R$ doublets, denoted respectively by $Q^{i }_{R}$ and $L^{i }_{R}$ ($i=1, \,2, \,3$). 
See Table \ref{table1} for the matter content of the model.

The scalar sector of the model features one $SU(2)_L$ doublet $H_L$ and one $SU(2)_R$ doublet $H_R$, which are respectively responsible for the EW symmetry and $SU(2)_R \times U(1)_{B-L}$ breaking. 
To obtain the realistic quark mass matrices with this scalar sector, we additionally introduce $SU(2)_L\times SU(2)_R$ singlet vectorlike quarks, $U^{i }_{L,R}$ and $D^{i }_{L,R}$ \cite{Balakrishna:1988bn, Babu:1988yq, Bolton:2019bou, Babu:2022ikf}. 
The quark sector is the same as the one in Refs.\,\cite{Bolton:2019bou,Babu:2022ikf}, except for global $Z^L_4 \times Z^R_4$ symmetries, which are imposed to forbid one-loop neutrino mass diagrams. 
To give heavy masses to the vectorlike quarks under these symmetry groups, an additional singlet Higgs field $S$ is added.

The lepton sector in our model is different from Refs.~\cite{Bolton:2019bou,Babu:2022ikf}.
The model contains neutral Dirac fermions $n^a_{L,R}$ (with $a=1,\cdots,N_f$), $SU(2)_L$ doublet scalars $\widetilde L^\alpha_{L}$ (with $\alpha=1,\cdots, N_s)$, and $SU(2)_R$ doublet scalars $\widetilde L^\alpha_{R}$. 
See Table \ref{table1} for the charge assignment. 
These extra fields play a crucial role in generating the realistic charged lepton and neutrino mass matrices (Sec.~\ref{sec:Me} and \ref{sec:Mnu}).
A global lepton number symmetry $U(1)_L$ is also imposed to keep neutrinos Dirac.
 
In this work we further invoke a LR exchange symmetry, which dictates that the total Lagrangian is invariant under the exchange between the left-handed and right-handed fields: 
\beq
Q^i_L \leftrightarrow Q^i_R,~U^i_L \leftrightarrow U^i_R,~D^i_L \leftrightarrow D^i_R, ~L^i_L \leftrightarrow L^i_R, ~n^a_L \leftrightarrow n^a_R,
\eeq
for the fermion fields, and 
\beq
S \leftrightarrow S^\dagger, ~H_L \leftrightarrow H_R, ~{\widetilde L}^{\alpha}_L \leftrightarrow {\widetilde L}^{\alpha}_R,
\eeq
for the scalar fields. 
The symmetry groups are also exchanged under this transformation: 
$SU(2)_L \leftrightarrow SU(2)_R$ and $Z^L_4 \leftrightarrow Z^R_4$. 
When this symmetry is extended to the spacetime parity symmetry, the model will provide a reasonable solution to the strong CP problem \cite{Babu:1989rb}. 

The scalar fields, $H_L$, $H_R$ and $S$, develop nonzero vacuum expectation values (VEVs)
\beq
\vev{H_L}=\begin{pmatrix} 0 \\ v_L \end{pmatrix},\quad
\vev{H_R}=\begin{pmatrix} 0 \\ v_R \end{pmatrix},\quad
\vev{S}=v_S\,,
\label{eq:VEVs}
\eeq
where $v_L \simeq 174$ GeV and all VEVs are real.
The LR symmetry is softly broken by scalar mass terms, which results in $v_L\neq v_R$. 
In this work, $v_R$ is assumed to be much larger than $v_L$.
The gauge symmetries are broken by these VEVs as follows:
\beq
SU(2)_R \times U(1)_{B-L} & \underset{\vev{H_R}}{\to} U(1)_Y \nonumber\\
SU(2)_L \times U(1)_Y & \underset{\vev{H_L}}{\to} U(1)_{\rm em}. \nonumber
\eeq
On the other hands, the scalar lepton fields $\wt{L}_{L,R}^\alpha$ are assumed not to develop the VEVs: $\vev{\wt{L}_L^\alpha}=\vev{\wt{L}_R^\alpha}=0$. 
Below, we discuss how fermion mass matrices are generated in our model.

\begin{table}[t]
\begin{center}
\begin{tabular}{|c|c|cccc|ccc|c}
\hline
Fields & spin   & ~$SU(3)_c$~ & ~$SU(2)_L$~  & ~$SU(2)_R$~  & ~$U(1)_{B-L}$~  &  ~$Z^L_4$ ~ &   ~$Z^R_4$ ~ & ~$U(1)_L$ ~    \\ \hline
  $Q^{i }_{L}$  & $1/2$ & ${\bf 3}$         &${\bf2}$    & ${\bf1}$    &            $1/3$    & $1$ & $1$  & $0$  \\
    $Q^{i }_{R}$   & $1/2$&  ${\bf 3}$        &${\bf1}$  &${\bf2}$      &            $1/3$         & $1$ & $1$  & $0$  \\ 
        $U^i_L$ & $1/2$ & ${\bf 3}$         &${\bf 1}$&${\bf1}$   &             $4/3$   & $1$ & $\omega$ &$0$   \\
          $U^i_R$ & $1/2$ & ${\bf 3}$         &${\bf 1}$&${\bf1}$   &             $4/3$   & $\omega$ & $1$ &$0$   \\ 
    $D^i_L$ & $1/2$ & ${\bf 3}$         &${\bf 1}$&${\bf1}$   &             $-2/3$   & 
 $1$ & $\omega^3$  &$0$   \\
          $D^i_R$ & $1/2$ & ${\bf 3}$         &${\bf 1}$&${\bf1}$   &             $-2/3$   & $\omega^3$ & $1$  &$0$   \\  
    \hline
     $L^{i }_{L}$ & $1/2$ & ${\bf 1}$         &${\bf2}$&${\bf1}$   &             $-1$    & $\omega$  & $1$ & $1$   \\
       $L^{i }_{R}$ & $1/2$ & ${\bf 1}$         &${\bf1}$&${\bf2}$   &             $-1$    & $1$ & $\omega$  & $1$  \\
    $n^a_L$ & $1/2$ & ${\bf 1}$         &${\bf 1}$&${\bf1}$   &             $0$ &  $1$  & $1$ &$1$   \\
          $n^a_R$ & $1/2$ & ${\bf 1}$         &${\bf 1}$&${\bf1}$   &             $0$  & $1$  & $1$  &$1$   \\   \hline \hline  
         $H_{L}$ & $0$ &  ${\bf 1}$        &${\bf2}$    &${\bf1}$   &       $1$      &$\omega$ & $1$   & $0$       \\  
        $H_{R}$ & $0$ &  ${\bf 1}$        &${\bf1}$    &${\bf2}$   &       $1$      & $1$ & $\omega$    & $0$       \\  \hline              
           $\widetilde L^\alpha_{L}$ & $0$ &  ${\bf 1}$        &${\bf2}$    &${\bf1}$   &       $-1$       &$\omega$  & $1$    & $0$       \\  
        $\widetilde L^\alpha_{R}$ & $0$ &  ${\bf 1}$        &${\bf1}$    &${\bf2}$   &       $-1$      & $1$  & $\omega$   & $0$       \\  
       $S$ & $0$ &  ${\bf 1}$        &${\bf1}$    &${\bf1}$   &       $0$      & $\omega^3$  & $\omega$    & $0$       \\            \hline
\end{tabular}
\end{center}
\caption{
Charge assignment of fermion and scalar fields in the model. 
$SU(3)_c\times SU(2)_L \times SU(2)_R \times U(1)_{B-L}$ are gauged, while 
$Z^L_4 \times Z^R_4 \times U(1)_L$ are global, where $\omega^4=1$ and $L$ is the lepton number.
$i=1, 2, 3$, $a=1, \cdots, N_f$ and $\alpha=1, \cdots, N_s$ denote flavor indices.
}
\label{table1}
\end{table}

\subsection{Quarks}
\label{sec:quark}

Quark masses are generated in a similar way to Refs.~\cite{Balakrishna:1988bn,Babu:1988yq}.  
The relevant couplings involving the colored fermions are 
\begin{eqnarray}
-{\cal L}_q&=&y^U_{ij}\, \ov{Q^{i }_{L}} \tau H^*_L  U^{j }_{R}+y^U_{ij}\, \ov{Q^{i }_{R}} \tau H^*_R  U^{j }_{L} + y^D_{ij}\, \ov{Q^{i }_{L}} H_L  D^{j }_{R}+y^D_{ij}\, \ov{Q^{i }_{R}} H_R  D^{j }_{L}  \nonumber   \\
&&  + \lambda^U_{ij}\, \ov{U^{i }_{L}} S U^{j }_{R} +\, \lambda^D_{ij}\, \ov{D^{i }_{L}} S^\dagger D^{j }_{R} + \hc,
\label{eq;quark}
\end{eqnarray}
where $\tau =i \sigma^2$. 
The LR symmetry dictates that all Yukawa couplings are $3 \times 3$ hermitian matrices. 
Assuming the vacuum alignment in Eq.\,(\ref{eq:VEVs}), we obtain the following mass terms, 
\beq
- {\cal L}_{{\cal M}_Q} 
    = \begin{pmatrix} \ov{\Hat u^i_L} & \ov{U^i_L} \end{pmatrix} 
        \begin{pmatrix} y^{U}_{ij} v_L & 0 \\ m^U_{ij} & y^{U *}_{ji} v_R \end{pmatrix}
        \begin{pmatrix}  U^j_R \\ \Hat u^j_R  \end{pmatrix}
    +\begin{pmatrix} \ov{\Hat d^i_L} & \ov{D^i_L} \end{pmatrix}
        \begin{pmatrix} y^{D}_{ij} v_L & 0 \\ m^D_{ij}  & y^{D *}_{ji} v_R \end{pmatrix}
        \begin{pmatrix}  D^j_R \\ \Hat d^j_R  \end{pmatrix},
\eeq
where $Q^{i }_{L}=(\Hat u^i_L \ \Hat d^i_L)^T$ and $Q^{i }_{R}=(\Hat u^i_R \ \Hat d^i_R)^T$, and $m^U_{ij}=\lambda^U_{ij} v_S$ and $m^D_{ij}=\lambda^D_{ij} v_S$ denote vectorlike quark masses.
Assuming the vectorlike quarks are heavy and integrating them out, we obtain three light colored fields for each quark type with their mass matrices,
\beq
\(M_u\)_{ij} \simeq v_L v_R \(y^U m_U^{-1} y^{U\dagger}\)_{ij} ,\quad
\(M_d\)_{ij} \simeq v_L v_R \(y^D m_D^{-1} y^{D\dagger}\)_{ij} .
\label{eq:Mq}
\eeq
These light fields are identified as the SM quarks. 
All SM fermion masses except the top quark can be explained using this approximate formula \cite{Balakrishna:1988bn, Babu:1988yq}. 

In discussion of neutrino mass generation, which we will study in Sec.~\ref{sec:Mnu}, 
mass mixing of the vectorlike quarks with the top and bottom quarks are important, so let us have a quick look at it here. 
For simplicity, we only consider one vectorlike quark in each up and down sector, denoted by $T_{L,R}:=U^3_{L,R}$ and $B_{L,R}:=D^3_{L,R}$, and ignore the other vectorlike quarks. 
The corresponding submatrices are given by
\beq
\begin{pmatrix} y^{U}_{33} v_L & 0 \\ m^U_{33}  & y^{U *}_{33} v_R  \end{pmatrix} &= U_{tL}  \begin{pmatrix} m_t &0 \\ 0 & m_T \end{pmatrix} U^\dagger_{tR}, \\
\begin{pmatrix} y^{D}_{33} v_L & 0 \\ m^D_{33}  & y^{D *}_{33} v_R  \end{pmatrix} &= U_{bL} \begin{pmatrix} m_b &0 \\ 0 & m_B \end{pmatrix} U^\dagger_{bR},
\eeq
where $U_{tL/tR}$ and $U_{bL/bR}$ are $2 \times 2$ unitary matrices, and $m_T$ and $m_B$ denote masses of the heavy quarks. 
For $|y_{33}^U| v_L \ll |m^U_{33}|, |y^U_{33}|v_R$, 
the heavy and light quark masses in the top sector are given by 
\beq
m_T \simeq \sqrt{\(m_{33}^U\)^2+\(y_{33}^U v_R\)^2} \,,\quad 
m_t \simeq \frac{|y_{33}^U|^2 v_L v_R}{m_T} \,.
\eeq
We see from this equation that $|y_{33}^U|v_R \ll |m_{33}^U|$ is not compatible with the observed top quark mass. 
In this work, we take $|y_{33}^U|v_R \gg m_{33}^U$ for simplicity, 
which yields $m_T \simeq |y_{33}^U| v_R$ and $m_t \simeq |y_{33}^U| v_L$. 
The mixing matrix elements are then given by 
$|\left(U_{tL}\right)_{11,22}|\simeq1$ and $|\left(U_{tL}\right)_{12,21}|\simeq\frac{m_t|m_{33}^U|}{m_T^2}$ for the left-handed quarks, and 
$|\left(U_{tR}\right)_{11,22}|\simeq1$ and $|\left(U_{tR}\right)_{12,21}|\simeq\frac{|m_{33}^U|}{m_T}$ 
for the right-handed quarks. 
The off-diagonal elements $|\left(U_{tR}\right)_{12,21}|$, which are important for neutrino mass generation, can be arbitrarily small by adjusting the ratio $|m_{33}^U|/m_T$.
The approximate equation (\ref{eq:Mq}) can be applied to the bottom sector. 
Assuming $|y_{33}^D| v_R \ll |m_{33}^D|$ for simplicity, 
we have $m_B \simeq m_{33}^D$ and $m_b \simeq \frac{|y_{33}^D|^2 v_L v_R}{m_{33}^D}$.
The mixing matrix elements are then given by 
$|\left(U_{bL}\right)_{11,22}| \simeq 1$ and $|\left(U_{bL}\right)_{12,21}| \simeq \frac{|y_{33}^D|v_L}{m_B}$ for the left-handed quarks, and 
$|\left(U_{bR}\right)_{11,22}| \simeq \frac{|y^D_{33}|v_R}{m_B}$ and 
$|\left(U_{bR}\right)_{12,21}| \simeq 1$ for the right-handed quarks.

\subsection{Charged leptons}
\label{sec:Me}

\begin{figure}[t]
\centering
\includegraphics[width=0.6\textwidth]{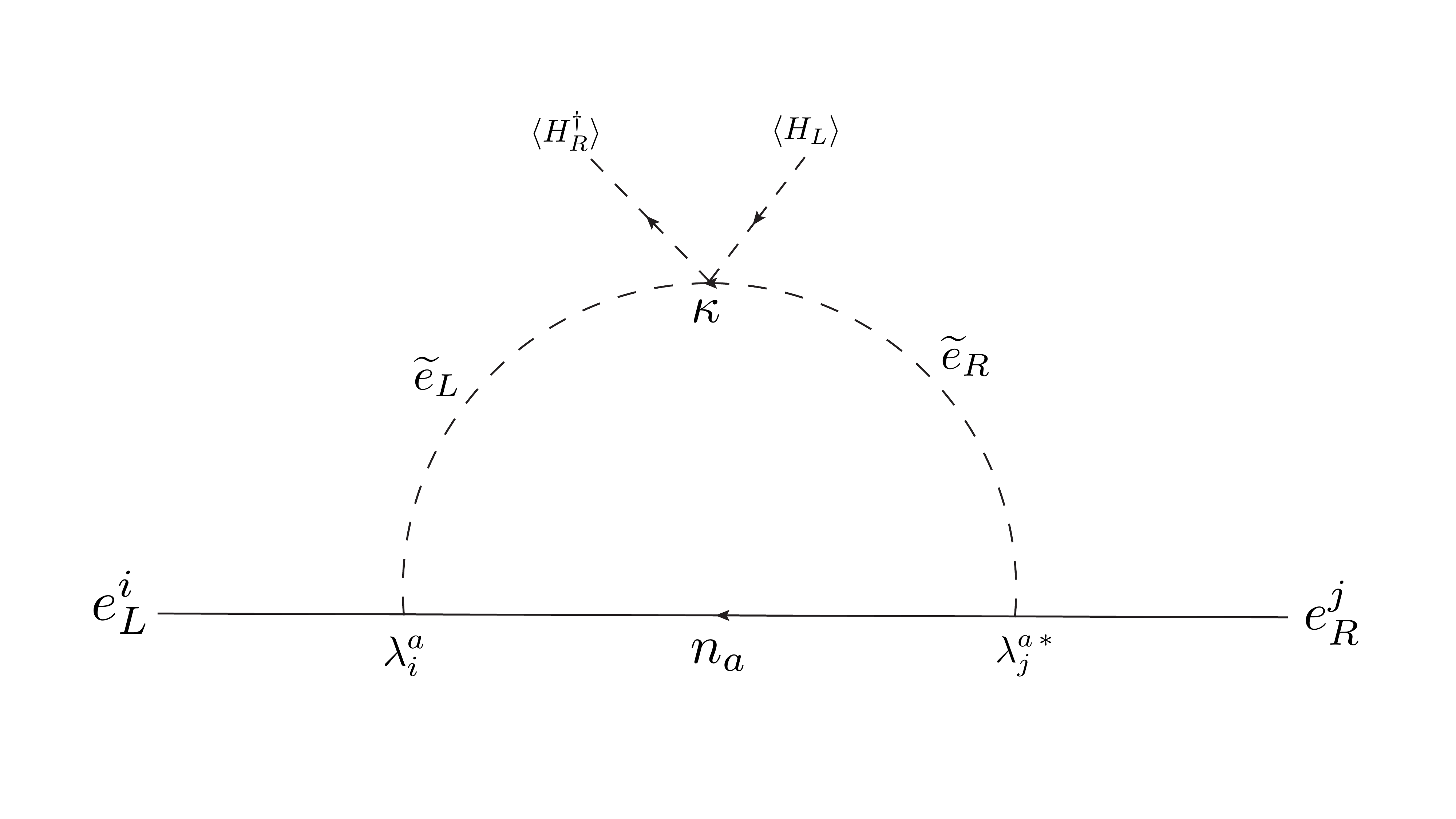}
\caption{Feynman diagram for the charged lepton mass matrix. 
}
\label{fig:one-loop_Me}
\end{figure}

In Refs.~\cite{Balakrishna:1988bn, Babu:1988yq, Babu:2022ikf, Bolton:2019bou}, 
the charged lepton masses are generated at the tree level in an analogous way to the quark masses. 
In our model, however, the doublet lepton fields $L_{L,R}^i$ do not directly couple to the doublet Higgs fields $H_{L,R}$, so their masses are generated in a different way. 
The key interactions are the couplings between 
$L_{L,R}^i$ and $\wt{L}_{L,R}^\alpha$, 
\begin{eqnarray}
\label{eq:L_portal}
M_a \,  \ov{n^a_L} n^a_R+\lambda^a_{i \alpha} \left (  \ov{ L^{i }_{L}} {\widetilde L}^{\alpha}_L n^a_{R}   + \ov{ L^{i }_{R}} {\widetilde L}^{\alpha}_R n^a_{L}  \right )+\hc 
\end{eqnarray}
and the couplings between $\widetilde L^\alpha_{L,R}$ and $H_{L,R}$, 
\beq
\kappa_{\alpha \beta} \left ( \widetilde L^{\alpha\, \dagger}_{L} H_L  \right ) \left (H^\dagger_R \widetilde L^{\beta}_{R}   \right ) + \hc
\eeq
The charged lepton masses are generated with these interactions at one-loop level  (Fig.~\ref{fig:one-loop_Me}). 
Assuming $\wt{L}_{L,R}^\alpha$ and $n_{L,R}^a$ are all heavy, this diagram induces the higher dimensional operator, 
\beq
\label{eq:Leff-Me}
-{\cal L}_{\eff, M_e} = \frac{y^{ij}_\ell}{v_R}  \ov{L^{i }_{L}}  \left ( H_L H^\dagger_R \right )L^{j}_{R} + \hc.
\eeq
This operator respects all gauge and global symmetries in our model and leads to the charged lepton masses when $H_L$ and $H_R$ develop the VEVs. 
Note that lepton chiral symmetries are broken only when $\lambda^a_{i\alpha}$, $M_a$ and $\kappa_{\alpha\beta}$ are all non-vanishing. 
The charged lepton masses are thus proportional to these couplings.

This is the basic mechanism to generate the charged lepton masses in our model. 
To fit both charged lepton masses and neutrino oscillation data, however, we need a non-trivial mass spectrum of $\wt{L}_{L,R}^\alpha$, which we will show in Sec.~\ref{sec:Mnu}. 
Such a non-trivial spectrum requires significant corrections from the VEVs of $H_{L,R}$, 
suggesting that we have to evaluate the one-loop charged lepton masses in the broken phase. 
We show the details below. 

Before calculating the one-loop charged lepton masses, it is convenient to tidy up the $\wt{L}_{L,R}^\alpha$ sector.
The general $\wt{L}_{L,R}^\alpha$ potential obeying the LR symmetry is given by
\begin{align}
V_{\wt{L}} 
    &=(m_L^2)_{\alpha\beta} \wt{L}_L^{\alpha\dagger} \wt{L}_L^\beta
	+ (m_R^2)_{\alpha\beta} \wt{L}_R^{\alpha\dagger} \wt{L}_R^\beta
	+ \left\{ \kappa_{\alpha\beta} \(\wt{L}_L^{\alpha\dagger} H_L\) \(H_R^\dagger \wt{L}_R^{\beta} \) + \hc \right\} \nonumber\\
    &\quad+\lam_{\alpha\beta}^e \left\{ \(\wt{L}_L^{\alpha\dagger} H_L\) \(H_L^\dagger \wt{L}_L^{\beta}\) 
	+ \(\wt{L}_R^{\alpha\dagger} H_R\) \(H_R^\dagger \wt{L}_R^{\beta}\) \right\} \nonumber\\
    &\quad+\lam_{\alpha\beta}^\nu \left\{ \(\wt{L}_L^{\alpha\dagger} \tau H_L^*\) \(H_L^T \tau^T \wt{L}_L^{\beta}\) 
	+ \(\wt{L}_R^{\alpha\dagger} \tau H_R^*\) \(H_R^T \tau^T \wt{L}_R^{\beta}\) \right\} \nonumber\\
    &\quad+\frac{1}{2} \lambda_{\alpha\beta}^s \left\{ \(\wt{L}_L^{\alpha\dagger} \tau H_L^*\) \(\wt{L}_L^{\beta\dagger} \tau H_L^*\) + \(\wt{L}_R^{\alpha\dagger} \tau H_R^*\) \(\wt{L}_R^{\beta\dagger} \tau H_R^*\) + \hc \right\} 
	+ \hat{V}_{\wt{L}} \,,
\end{align}
where $\lambda^s_{\alpha\beta} = \lambda^s_{\beta\alpha}$, and 
$\hat{V}_{\wt{L}}$ contains all quartic terms that do not contribute to the $\wt{L}_{L,R}^\alpha$ masses. 
Here, $\kappa_{\alpha \beta}$, $(m^2_{L/R})_{\alpha \beta}$, $\lambda^e_{\alpha \beta}$ and $\lambda^\nu_{\alpha \beta}$ are $N_s \times N_s$ hermitian matrices, and the LR symmetry is assumed to be softly broken, i.e. $\left(m^2_{L} \right)_{\alpha \beta}\neq \left(m^2_{R} \right)_{\alpha \beta}$. 
The origin of the softly breaking is beyond our scope.

$\wt{L}_{L,R}^\alpha$ are decomposed into the neutral and charged components: 
\beq
\wt{L}_L^\alpha = \begin{pmatrix} \wt{N}_L^\alpha \\ \wt{E}_L^\alpha \end{pmatrix} \,,\quad
\wt{L}_R^\alpha = \begin{pmatrix} \wt{N}_R^\alpha \\ \wt{E}_R^\alpha \end{pmatrix} \,.
\eeq
After the symmetry breaking, we have the mass terms, 
\begin{align}
V_{\wt{L}} 
    &\supset \( (m_L^2)_\alpha \delta_{\alpha\beta} + \lam_{\alpha\beta}^e v_L^2 \) \wt{E}_L^{\alpha\dagger} \wt{E}_L^\beta 
	+ \( (m_R^2)_\alpha \delta_{\alpha\beta} + \lam_{\alpha\beta}^e v_R^2 \) \wt{E}_R^{\alpha\dagger} \wt{E}_R^\beta \nonumber\\
    &\quad + \(\kappa_{\alpha\beta} v_L v_R \, \wt{E}_L^{\alpha\dagger} \wt{E}_R^\beta + \hc\) \nonumber\\
    &\quad + \( (m_L^2)_\alpha \delta_{\alpha\beta} + \lam_{\alpha\beta}^\nu v_L^2 \) \wt{N}_L^{\alpha\dagger} \wt{N}_L^\beta 
		+ \frac{1}{2} \lambda^s_{\alpha\beta} v_L^2 \(\wt{N}_L^\alpha \wt{N}_L^\beta + \hc\) \nonumber\\
    &\quad + \( (m_R^2)_\alpha \delta_{\alpha\beta} + \lam_{\alpha\beta}^\nu v_R^2 \) \wt{N}_R^{\alpha\dagger} \wt{N}_R^\beta 
		+ \frac{1}{2} \lambda^s_{\alpha\beta} v_R^2 \(\wt{N}_R^\alpha \wt{N}_R^\beta + \hc\) \,.
\end{align}
It is easy to see that the mass differences between the charged and neutral states are generated by the VEVs. 
The charged scalar mass terms are given in the form 
\beq
V_{{\cal M}_{\wt{e}}} = 
    \begin{pmatrix} 
        \wt{E}_L^{\alpha \dagger} & \wt{E}_R^{\gamma \dagger} 
    \end{pmatrix}
    \underbrace{
    \begin{pmatrix}
        (\hat{m}_{\wt eL}^2)_{\alpha\beta} & \kappa_{\alpha\delta} v_L v_R \\ 
	(\kappa^\dagger)_{\gamma\beta} v_L v_R & (\hat{m}_{\wt eR}^2)_{\gamma\delta}
    \end{pmatrix}
    }_{\equiv {\cal M}_{\wt{e}}^2}
    \begin{pmatrix} 
	\wt{E}_L^{\beta} \\ \wt{E}_R^{\delta} 
    \end{pmatrix} \,,
\eeq
where 
\beq
(\hat{m}_{\wt eL}^2)_{\alpha\beta} = (m_L)_\alpha^2 \delta_{\alpha\beta} + \lambda^e_{\alpha\beta} v_L^2 \,,\quad
(\hat{m}_{\wt eR}^2)_{\alpha\beta} = (m_R)_\alpha^2 \delta_{\alpha\beta} + \lambda^e_{\alpha\beta} v_R^2 \,.
\eeq
The $\kappa$ terms induce the left-right scalar mixing. 
We expect that without any fine-tuning, the diagonal blocks of ${\cal M}_{\wt e}^2$ are characterized by the scales of the VEVs, 
\beq
\hat{m}_{\wt eL}^2 \sim {\cal O}(v_L^2) \,,\quad
\hat{m}_{\wt eR}^2 \sim {\cal O}(v_R^2) \,.
\eeq
In this case, the left-right mixing is small unless $\kappa$ is extremely large. 
We here consider the case where $\kappa$ is ${\cal O}(1)$ or smaller, allowing to approximately diagonalize ${\cal M}_{\wt e}^2$ by a unitary matrix $U_{\wt e}$, 
\beq
{\cal M}_{\wt{e}}^2 = U_{\wt{e}} \, {\cal M}_{\wt{e}, {\rm diag}}^2 \, U_{\wt{e}}^\dagger, \quad 
U_{\wt{e}} \simeq \begin{pmatrix} V_{\wt{e}_L} & 0 \\ 0 & V_{\wt{e}_R} \end{pmatrix} \begin{pmatrix} \1 & x_{LR} \\ - x_{LR}^\dagger & \1 \end{pmatrix} \,.
\eeq
Here, $V_{\wt{e}_L}$ ($V_{\wt{e}_R}$) are unitary matrices that diagonalize $\hat{m}_{\wt eL}^2$ ($\hat{m}_{\wt eR}^2$), 
\beq
\hat{m}_{\wt eL}^2 = V_{\wt e_L}  m_{\wt eL, {\rm diag}}^2 V_{\wt e_L}^\dagger \,,\quad
\hat{m}_{\wt eR}^2 = V_{\wt e_R}  m_{\wt eR, {\rm diag}}^2 V_{\wt e_R}^\dagger \,,
\eeq
where $(m_{\wt eL, {\rm diag}}^2)_{\alpha\beta}=m_{\wt eL\alpha}^2 \delta_{\alpha\beta}$ and $(m_{\wt eR, {\rm diag}}^2)_{\alpha\beta}=m_{\wt eR\alpha}^2 \delta_{\alpha\beta}$. 
The mass eigenstates $\widetilde e^\alpha_{L/R}$ are then given by
\begin{equation}
\wt E^\alpha_{L/R} = \left(V_{\wt e_L/\wt e_R}\right)_{\alpha\beta} \wt e^\beta_{L/R}.
\end{equation}
The concrete expression of $x_{LR}$ can be obtained by solving the Sylvester equation,
\beq
m_{\wt eL}^2 x_{LR} - x_{LR} m_{\wt eR}^2 = - v_L v_R V_{\wt e_L}^\dagger \kappa V_{\wt e_R} \,.
\eeq
For $m_{\wt eL}^2 \ll m_{\wt eR}^2$, we can ignore the first term in the left-hand side, finding 
\beq
(x_{LR})_{\alpha\beta} \simeq \frac{v_L v_R}{m_{\wt eR\beta}^2} (V_{\wt{e}_L}^\dagger \kappa V_{\wt{e}_R})_{\alpha\beta} \,.
\eeq

Now we are ready to resume the one-loop mass generation of the charged leptons. 
First, we diagonalize the mass matrices of $\wt{E}_{L,R}^\alpha$. 
Then the relevant interaction Lagrangian is given by 
\begin{align}
- {\cal L}_L  =& (\lambda_L^a)_{i\alpha} \ol{e^i_L} \wt{e}^\alpha_L n^a_R + (\lambda_R^a)_{i\alpha} \ol{e^i_R} \wt{e}^\alpha_R n^a_L + M_a \, \ol{n_L^a} n_R^a + \hc \\
 & + m_{\wt eL\alpha}^2 \wt{e}_L^{\alpha\dagger} \wt{e}_L^\alpha + m_{\wt eR\alpha}^2 \wt{e}_R^{\alpha\dagger} \wt{e}_R^\alpha
		+ \(\wt{\kappa}_{\alpha\beta} v_L v_R \, \wt{e}_L^{\alpha\dagger} \wt{e}_R^\beta + \hc\),
\label{eq:DM_int2}
\end{align}
where 
\beq
(\lambda_L)^a_{i\alpha} = (\lambda^a V_{\wt{e}_L})_{i\alpha} \,,\quad
(\lambda_R)^a_{i\alpha} = (\lambda^a V_{\wt{e}_R})_{i\alpha} \,,\quad
\wt{\kappa}_{\alpha\beta} = (V_{\wt{e}_L}^\dagger \kappa V_{\wt{e}_R})_{\alpha\beta} \,.
\eeq
The one-loop diagram in Fig.~\ref{fig:one-loop_Me} can be analytically calculated, yielding the charged lepton mass terms, 
\beq
\label{eq:lepton_mass}
-{\cal L}_{\eff, M_e} = \sum_a (m_e)^a_{ij} \, \ol{e_L^i} e_R^j + \hc,
\eeq
where 
\beq
(m_e)^a_{ij} = \sum_{\alpha,\beta} (\lambda_L)^a_{i\alpha} \wt{\kappa}_{\alpha\beta} (\lambda_R^*)^a_{j\beta}\frac{v_Lv_R M_a}{(4 \pi)^2}
F(M_a^2,m^2_{\wt{e}L\alpha},m^2_{\wt{e}R\beta}).
\eeq
Here, the loop function is given by
\begin{eqnarray}
F(M_a^2,m^2_{\widetilde e L \alpha}, m^2_{\widetilde e R \beta})
&=& \frac{ m^2_{\widetilde e L \alpha}   \log \left ( \frac{m^2_{\widetilde e L \alpha}}{M_a^2} \right ) }{\left ( m^2_{\widetilde e L \alpha} -M_a^2 \right ) \left ( m^2_{\widetilde e L \alpha} - m^2_{\widetilde e R \beta} \right )  } -\frac{ m^2_{\widetilde e R \beta}   \log \left ( \frac{m^2_{\widetilde e R \beta}}{M_a^2} \right ) }{\left ( m^2_{\widetilde e R \beta} -M_a^2 \right ) \left ( m^2_{\widetilde e L \alpha} - m^2_{\widetilde e R \beta} \right )  }. \nonumber \\
\label{eq:loop_function}
\end{eqnarray}
We evaluated the one-loop diagram using the $\kappa$-insertion approximation, but the full calculation gives the same result as long as $\kappa$ is small. 
This approximation is sufficient in our discussion.

\subsection{Neutrinos}
\label{sec:Mnu}

\begin{figure}[t]
\centering
\includegraphics[width=8cm]{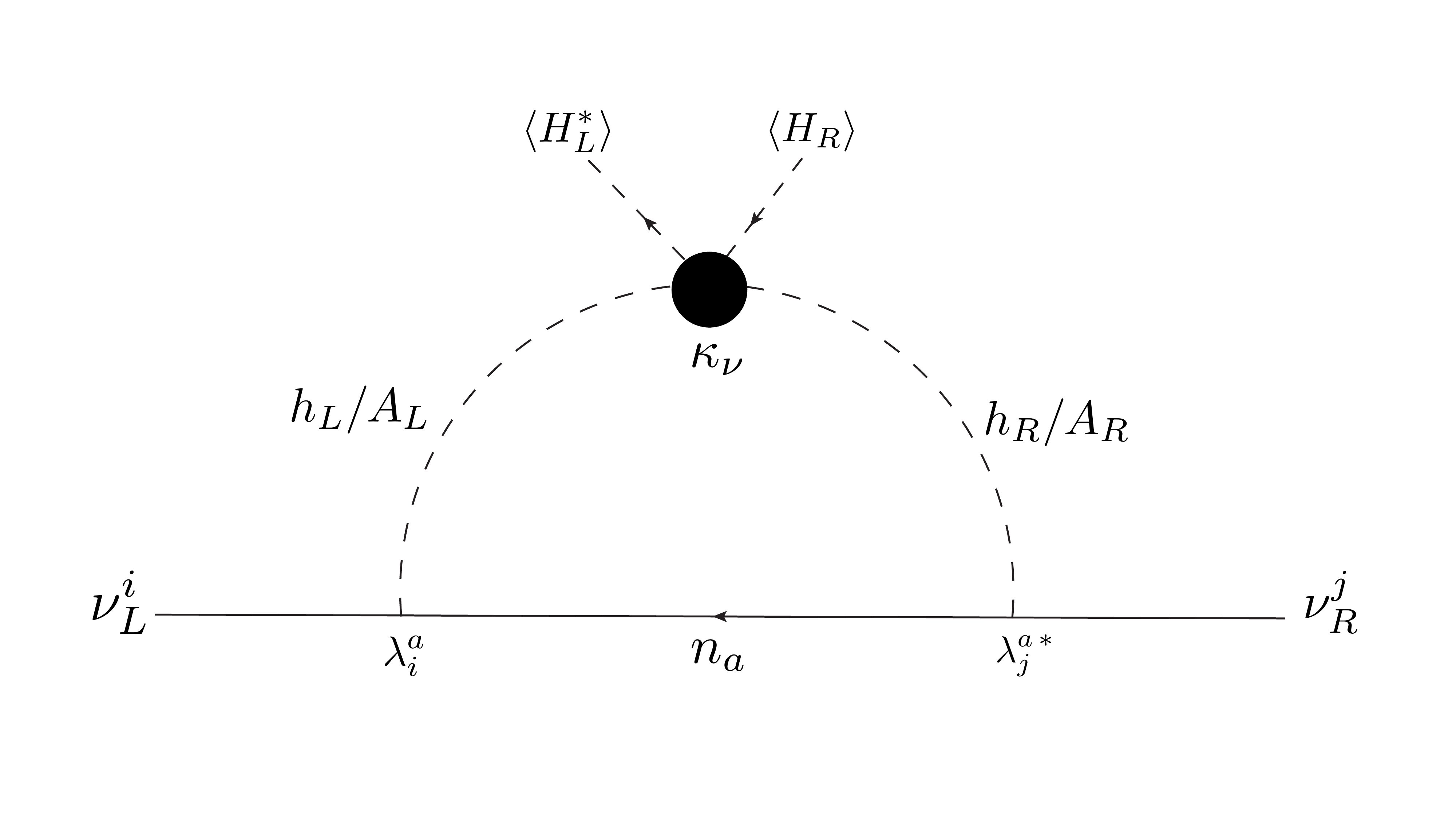}
\caption{Feynman diagram for the neutrino mass matrix. 
}\label{fig:one-loop_Mnu}
\end{figure}

The $U(1)_L$ global symmetry forbids Majorana mass terms of the neutral fermions. 
As a result, the neutrinos are Dirac fermions in our model.
We naively expect that, similarly to the charged lepton, the neutrino mass terms are generated from the effective operator like 
\begin{eqnarray}
-{\cal L}_{\eff,M_\nu} &=& \frac{y^{ij}_\nu}{v_R}  \ov{L^{i }_{L}}  \left ( \tau H^*_L H^T_R \tau^T \right )L^{j}_{R} + \hc, 
\label{eq:Leff_Mnu}
\end{eqnarray}
which is an analog to Eq.\,(\ref{eq:Leff-Me}). 
This operator could be induced via the one-loop diagram shown in Fig.\,\ref{fig:one-loop_Mnu} if there were a new quartic interaction, 
\beq
\left(\kappa_\nu\right)_{\alpha\beta} \left(\widetilde L^{\alpha\dagger}_{L} \tau H^*_L \right) \left(H^T_R \tau^T  \widetilde L^\beta_{R}\right) + \hc,
\label{eq:kappa_nu}
\eeq
when combined with Eq.\,(\ref{eq:L_portal}).
However, this term breaks the $Z^L_4 \times Z^R_4$ symmetry and does not exist in the tree level Lagrangian, so the neutrinos do not obtain their masses at the one-loop level. 
In fact, nonzero neutrino masses are generated at higher-loop levels, due to an effect similar to what is caused by the $\kappa_\nu$ term. 
That effect arises from picking up the $Z^L_4 \times Z^R_4$ breaking in the loop. 

\begin{figure}[t]
\centering
\includegraphics[width=7cm]{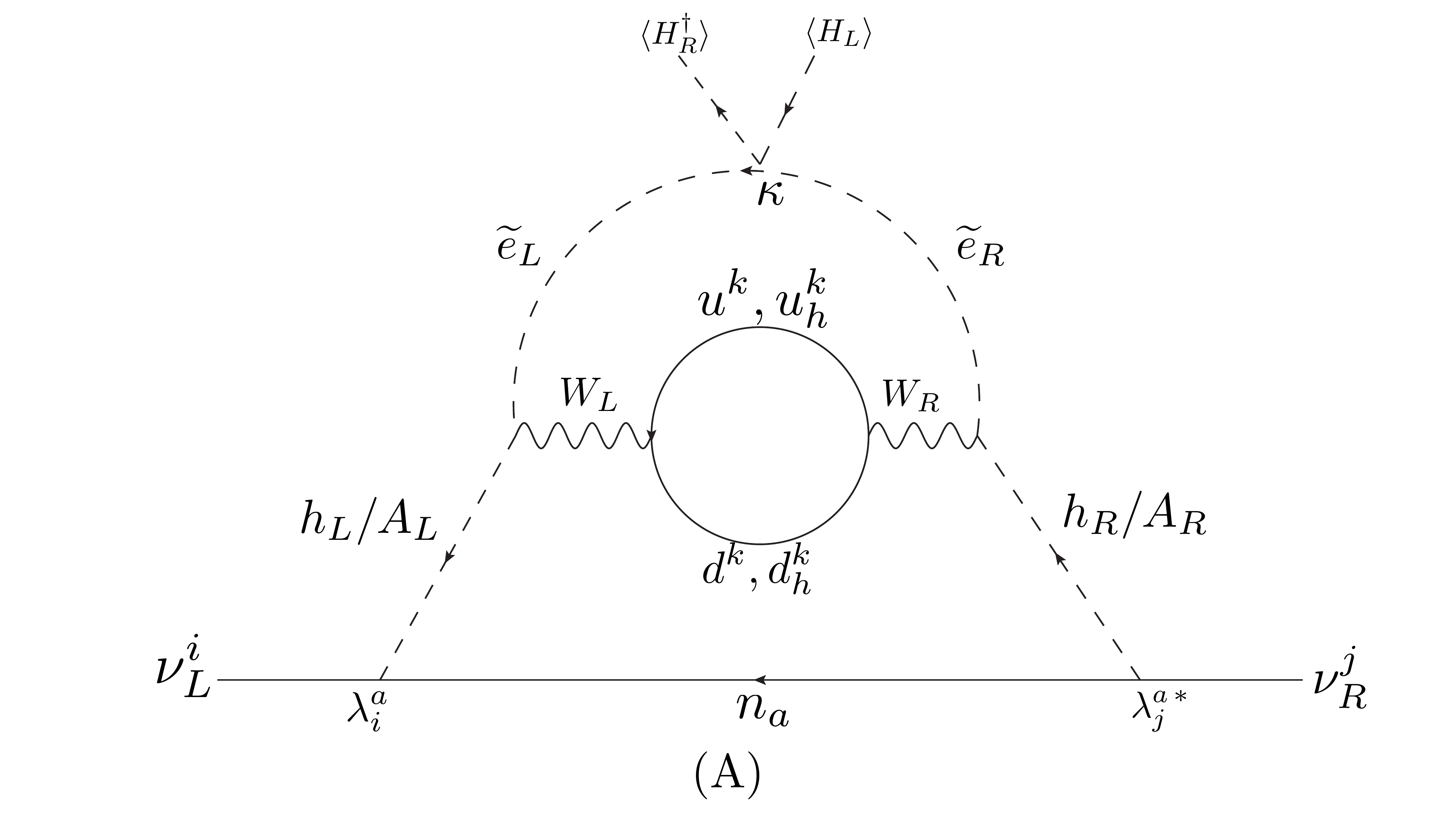}\includegraphics[width=7cm]{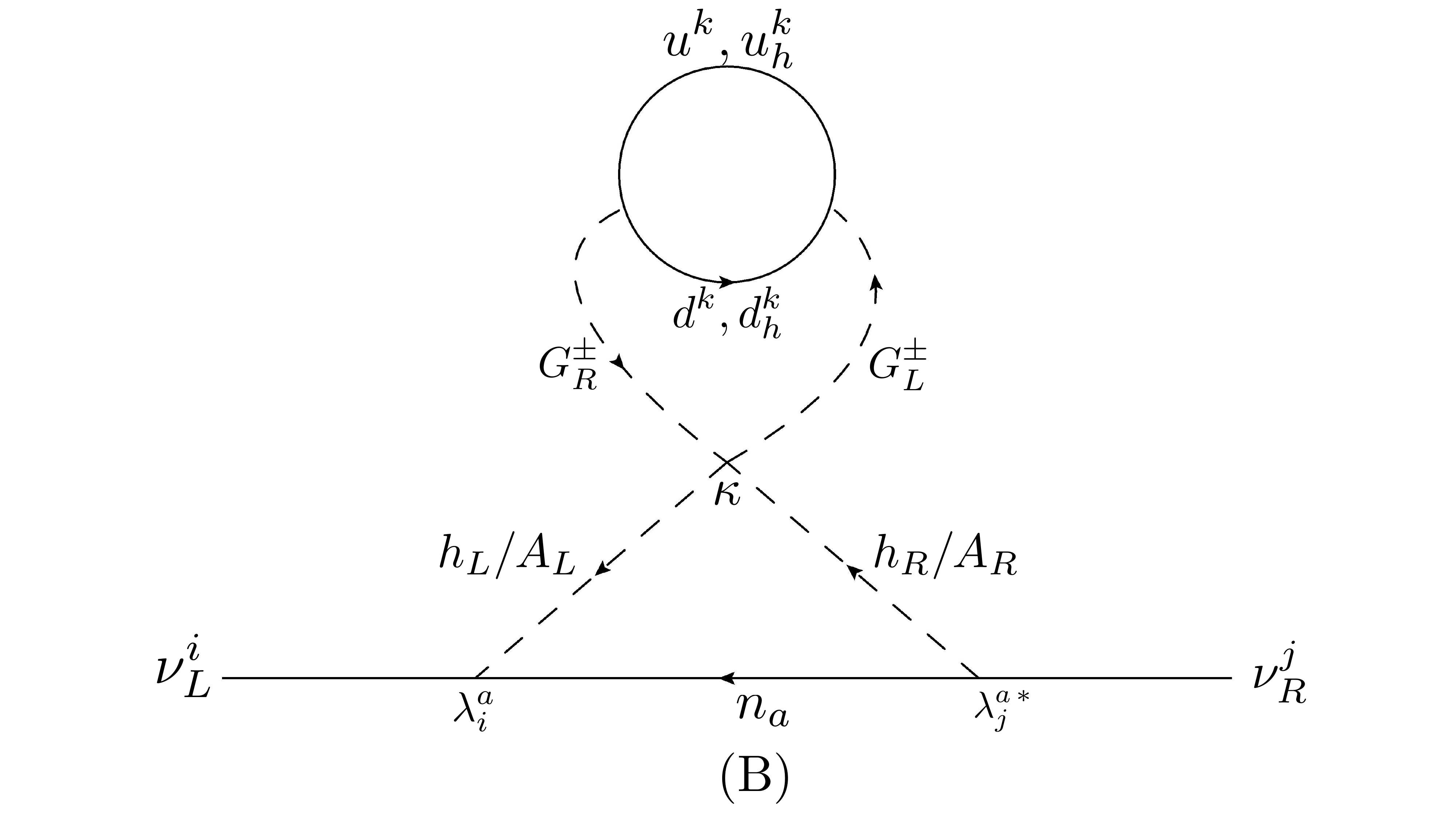}
\includegraphics[width=7cm]{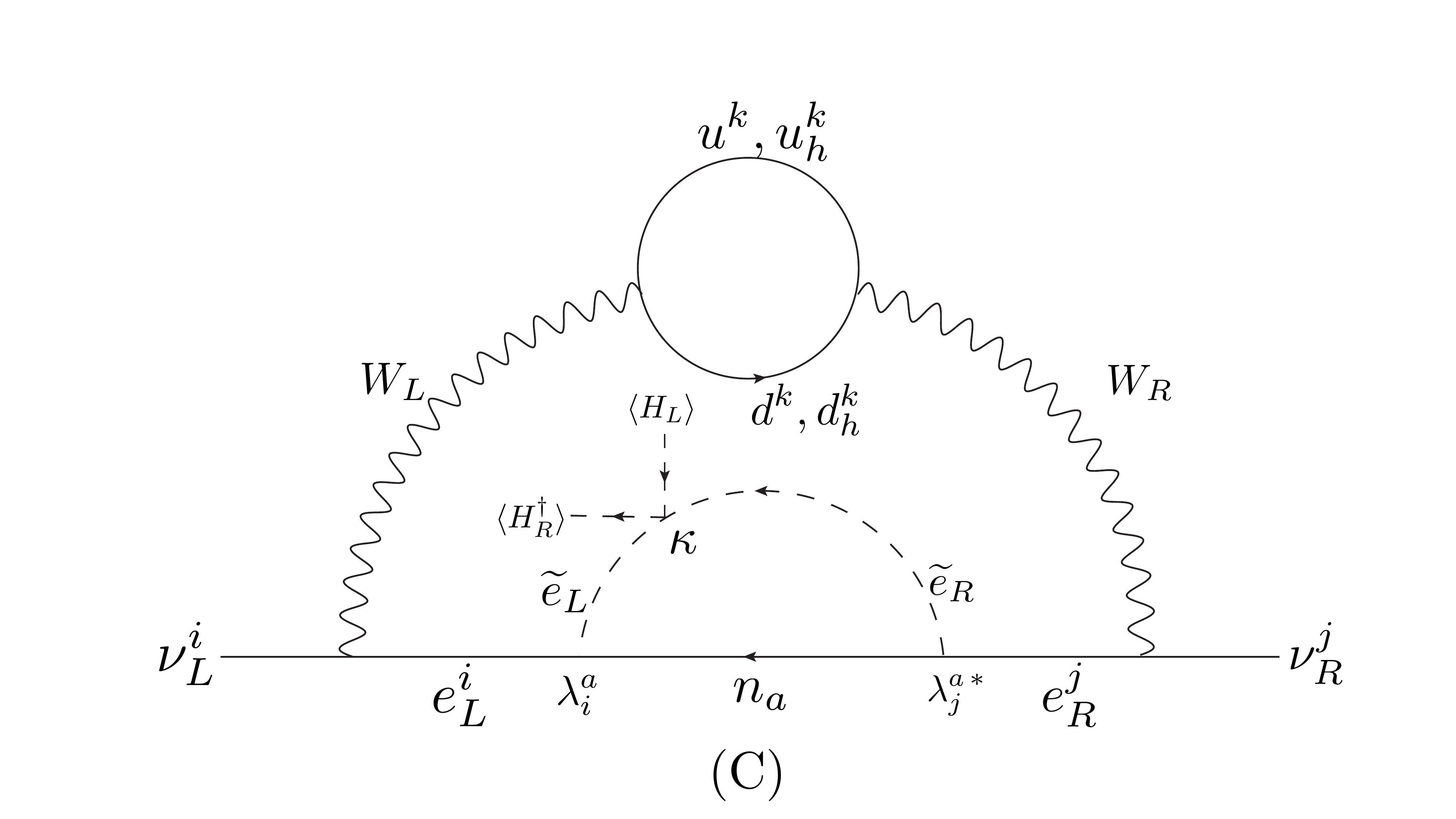}
\caption{Feynman diagrams for three-loop neutrino mass generation. 
The light and heavy up-type (down-type) quarks are denoted by $u^k$ and $u^k_h$ ($d^k$ and $d^k_h$).
}\label{fig:three-loop}
\end{figure}

A crucial role of the $\kappa_\nu$ term in generating the neutrino masses is left-right mixing of the neutral scalar leptons $\wt{N}_{L,R}$:
\begin{equation}
\left (m^2_{LR} \right )_{\alpha \beta}\,\wt{N}_L^{\alpha\dagger} \wt{N}_R^\beta+\hc
\end{equation}
In our model, such mass mixing can be generated from two-loop diagrams involving the top and bottom quarks, so the neutrino masses are generated at the three-loop order.
See the diagrams (A) and (B) in Fig.~\ref{fig:three-loop}, where one can find the two-loop left-right neutral scalar mixing is contained as subdiagrams. 
For convenience, we here define the {\it equivalent} $\kappa_\nu$ parameter by $(\kappa_\nu^{\rm equiv})_{\alpha\beta}:=\left(m^2_{LR}\right)_{\alpha\beta}/\left(v_Lv_R \right)$. 
Once we obtain the left-right mass mixing,
the induced neutrino mass matrix can be calculated as
\beq
\label{eq:L_Mnu}
-{\cal L}_{M_\nu} &= (\Hat M_\nu)_{ij} \,\ov{\nu^{i }_{L}} \nu^{j}_{R} + \hc ,\nonumber\\
(\Hat M_\nu)_{ij} & = v_L v_R \left\{\sum^{N_f}_{a=1} \sum_{\alpha,\beta} \sum_{S_L,S_R} \frac{(\lambda_{S_L})^{a}_{i\alpha} \left(\kappa_\nu^{S_L S_R}\right)_{\alpha\beta} (\lambda_{S_R}^{*})^{a}_{j\beta}}{4\,(4 \pi)^2} 
\, M_a F(M_a^2, m^2_{S_L\alpha}, m^2_{S_R\beta})\right\},
\eeq
where $S_L=h_L, A_L$ and $S_R=h_R, A_R$ are the component fields of $\wt N_{L,R}$: 
\beq
\widetilde N^{\alpha }_{L/R}
&=\frac{1}{\sqrt{2}}\({\Hat h}^\alpha_{L/R} + i {\Hat A}^\alpha_{L/R}\).
\eeq
To evaluate Eq.~(\ref{eq:L_Mnu}), 
we diagonalized the neutral scalar mass matrices $\Hat m^2_{S_L/S_R}$ using unitary matrices, 
\beq
\Hat m^2_{S_L}=V_{S_L} m^2_{S_L, {\rm diag}} V^\dagger_{S_L},\quad
\Hat m^2_{S_R}=&V_{S_R} m^2_{S_R, {\rm diag}} V^\dagger_{S_R},
\eeq
where $(m^2_{S_L, {\rm diag}})_{\alpha\beta}=m^2_{S_L\alpha}\delta_{\alpha\beta}$ and $(m^2_{S_R, {\rm diag}})_{\alpha\beta}=m^2_{S_R\alpha}\delta_{\alpha\beta}$.
We also defined the mass eigenstates, $h^\alpha_{L/R}$ and $A^\alpha_{L/R}$, by 
\begin{equation}
\Hat S_L^\alpha = \left(V_{S_L}\right)_{\alpha\beta}  S_L^\beta,\quad
\Hat S_R^\alpha = \left(V_{S_R}\right)_{\alpha\beta} S_R^\beta,
\end{equation}
and their couplings by 
\beq
\label{Yukawa-neutral}
\(\lambda_{S_L}\)^{a}_{i\alpha} = \left(\lambda^a V_{S_L}\right)_{i \alpha},\quad
\(\lambda_{S_R}\)^{a}_{i\alpha} = \left(\lambda^a V_{S_R}\right)_{i\alpha},\quad
\left(\kappa^{S_L S_R}_\nu \right)_{\alpha\beta} = (V^\dagger_{S_L} \kappa_\nu^{\rm equiv} V_{S_R})_{\alpha \beta}.
\eeq
$\kappa_\nu^{\rm equiv}$ induced from the diagrams (A) and (B) is calculated in Appendix \ref{appendix}. 
Note that we calculated $(\hat M_\nu)_{ij}$ in the $\kappa$-insertion scheme again. 

It is worth mentioning that 
the flavor structure of $(\Hat M_\nu)_{ij}$ tends to align with the charged lepton mass matrix. 
This alignment precisely happens when the neutral scalars $S_{L,R}$ have the similar mass spectrum and mixing pattern to those of the charged scalars $\wt e_{L,R}$. 
Since the neutrino mass structure is quite different from the charged lepton one in nature, we need to accommodate different structures with the neutral and charged scalars. 
In this work, we introduce a mass hierarchy only in the neutral scalar sector, which we we will discuss in detail in Sec.~\ref{sec:illustration}.

Note that the neutrino mass matrix $\Hat M_\nu$ receives additional contribution directly from other three-loop diagram, not through the $\wt{N}_{L,R}$ mixing. 
See the diagram (C) in Fig.~\ref{fig:three-loop}.  
This contribution is similar to the one studied in Refs. \cite{Balakrishna:1988bn,Babu:1988yq,Babu:2022ikf,Bolton:2019bou}. 
Given that the charged scalars $\wt e^\alpha_{L/R}$ are much heavier than the SM leptons, we can approximate the inner subdiagram to the one-loop induced charged lepton mass matrix in evaluating the neutrino masses. 
Then, the diagram (C) can be evaluated approximately as a two-loop diagram, with a mass insertion in the charged lepton propagator.
As a result, the diagram (C) contribution to the neutrino mass matrix precisely aligns with the charged lepton one: $(\Hat M^{(C)}_\nu)_{ij}\propto\left(m_e \right)_{ij}$. 
This contradicts with the observed neutrino mixing, if it dominates the neutrino mass generation. Since this contribution is smaller than the other two in parameter space of our interest, we will not consider this contribution further.

Lastly, let us give a brief comment on a potential UV source of $\kappa_\nu$.
In the presence of additional heavy fields charged under $Z_4^L \times Z_4^R$, the $\kappa_\nu$ term might be effectively generated. 
For example, one can write down a dim-6 operator like 
\beq
{\cal L} \supset \frac{(c_\nu)_{\alpha\beta}}{\Lambda^2} S^2 \left(\widetilde L^{\alpha\dagger}_{L} \tau H^*_L \right) \left(H^T_R \tau^T  \widetilde L^\beta_{R}\right) + \hc,
\label{eq:kappa_nu_UV}
\eeq
which could be obtained after integrating out those unspecified heavy fields. 
This term yields an effective $\kappa_\nu$ term, after $S$ develops the VEV:
\beq
\left (\kappa^{\rm UV}_\nu \right )_{\alpha \beta} = \left (c_\nu \right )_{\alpha \beta}  \frac{v_S^2}{\Lambda}.
\eeq
In this case, the suppression of $\kappa_\nu$ might be understood by a large scale separation between $\Lambda$ and $v_S$, where $\Lambda$ would be related to mass of the additional heavy states. 
In the next section, we analyze the observables in the lepton sector assuming $\kappa_\nu^{\rm equiv}$ is the only source of $\kappa_\nu$.
When some parameters cannot precisely be fit to the observed quantities, one should expect that $\kappa^{\rm UV}_\nu $ will compensate for the deviations.

\section{Fitting lepton masses and mixing}
\label{sec:illustration}

We have outlined our idea for generating the fermion masses in the LR symmetric model. 
However, this mechanism is viable only in specific regions of parameter space, because in general it leads sizable charged lepton flavor violation (cLFV) and an undesired alignment of the neutrino mass matrix with the charged lepton one. 
Thus, we need to consider a non-trivial flavor structure for the new particles. 
In this section, we demonstrate that assuming a flavor structure for the new neutral scalars, $h_{L,R}$ and $A_{L,R}$, we can achieve the observed patterns of the lepton mass matrices.

For concreteness, we take $N_s=N_f=3$, which is arguably a minimal setup to realize the observed mass matrix structure in the lepton sector. 
We further assume 
\begin{equation}
\lambda^a_{i\alpha} = \lambda^a_i\,\delta_{i\alpha}\,,\quad
\lambda^e_{\alpha\beta} = \lambda^e_\alpha\,\delta_{\alpha\beta}\,,\quad
\kappa_{\alpha\beta} = \kappa_\alpha\,\delta_{\alpha\beta} \,.
\label{eq:assumption1}
\end{equation}
This assumption guarantees no mass mixing among the charged scalar leptons at least at the one-loop level,
thereby suppressing cLFV processes. 
Nevertheless, the neutrino mass matrix can be correctly generated in the presence of the mass mixing among the neutral scalars, as we will see below. 
To further simplify the analysis,
we also consider the universal charged scalar masses:
\beq
m_{\wt eL\alpha} =: m_{\wt eL} \,,\quad
m_{\wt eR\alpha} =: m_{\wt eR} \,.
\label{eq:assumption2}
\eeq
The survey of the entire parameter space by scanning these charged scalar masses and other parameters is beyond the scope of this paper. 
In the following, we will try to derive the observed lepton mass structures from these couplings.

\subsection{Charged lepton masses}

\begin{figure}[t]
    \centering
    \includegraphics[width=10cm]{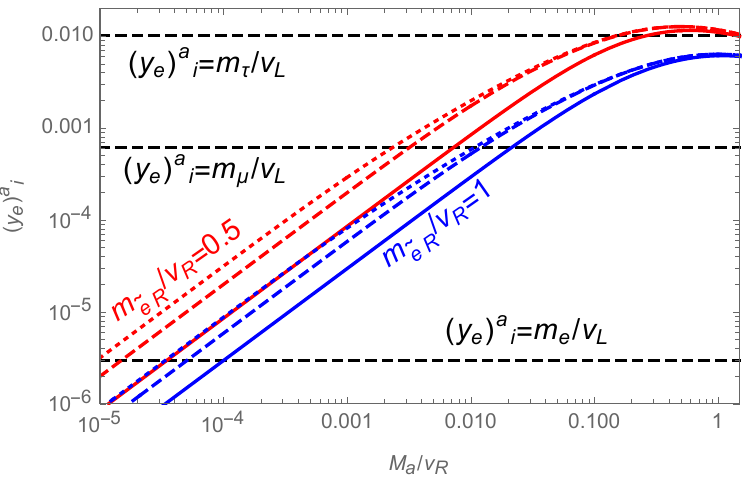}
    \caption{The $n^a$ contribution to the effective Yukawa coupling of a charged lepton $e_i$, defined by $(y_e)^{a}_{i}:=(m_e)^{a}_{i}/v_L$, which we calculate using Eq.~(\ref{eq:charged}). 
    We take $\kappa_{i}|\lambda^{a}_{i}|^2=1$ and assume the universal mass spectrum for the charged scalars, $m_{\wt e L \alpha}=:m_{\wt e L}$ and $m_{\wt e R \alpha}=:m_{\wt e R}$. 
    The red (blue) line corresponds to $m_{\wt e R}/v_R=1$ ($0.5$), with $m_{\widetilde e L}/v_R=0.1$ (solid), $0.01$ (dashed) and $0.001$ (dotted). 
    The black horizontal dashed lines represent the reference values of the realistic charged lepton Yukawa couplings, $(y_e)^a_i=m_e/v_L$, $m_\mu/v_L$, and $m_\tau/v_L$.
    }\label{fig:charged}
\end{figure}

Under two assumptions (\ref{eq:assumption1})-(\ref{eq:assumption2}), 
the charged lepton mass matrix $(m_e)^{a}_{ij}$ becomes diagonal: 
\beq
(m_e)^{a}_{ij} = (m_e)^a_i\,\delta_{ij} 
= \frac{\kappa_i|\lam^a_i|^2}{(4\pi)^2} 
v_L v_R M_a F(M_a^2,m^2_{\wt{e}L}, m^2_{\wt{e}R} ) \, \delta_{ij}.
\label{eq:charged}
\eeq
In Fig.~\ref{fig:charged}, 
we show the $n^a$ contribution to the effective Yukawa coupling of a charged lepton $e_i$, defined by $(y_e)^{a}_{i}:=(m_e)^{a}_{i}/v_L$. 
We take $\kappa_{i} |\lambda^{a}_{i}|^2=1$ for reference. 
The red (blue) line corresponds to $m_{\wt e R}/v_R=1$ ($0.5$), with $m_{\wt e L}/v_R=0.1$ (solid), $0.01$ (dashed) and $0.001$ (dotted), respectively. 
The black dashed lines represent the reference values of the realistic charged lepton Yukawa couplings, $(y_e)^a_i=m_e/v_L$, $m_\mu/v_L$, and $m_\tau/v_L$. 
We find $M_a/v_R \lesssim {\cal O}(10^{-4})$ for the electron, $\lesssim {\cal O}(10^{-2})$ for the muon and $\lesssim {\cal O}(1)$ for the tau with $\kappa_{i} |\lambda^{a}_{i}|^2=1$. 
Either a light $n^a$ or small $\lambda^{a}_{i}$ is thus required to explain the light charged lepton masses.

To see how the charged lepton spectrum is reproduced in our model, 
it is useful to analytically approximate Eq.~(\ref{eq:charged}). 
For $m_{\wt{e}L}, M_a \ll m_{\wt{e}R}$, 
we approximately obtain the effective lepton Yukawa couplings, 
\beq
(y_e)_{i} := \sum_a \frac{(m_e)^a_i}{v_L} \simeq \sum_a \frac{\kappa_i |\lambda_i^a|^2}{(4\pi)^2} \frac{v_R}{M_a} \frac{M_a^2}{m_{\wt{e}R}^2} \log\frac{m_{\wt{e}R}^2}{M_a^2} \,.
\label{eq:ye_approx}
\eeq
To reproduce the tau yukawa coupling $y_\tau \sim 10^{-2}$, we have to require 
\beq
\frac{v_R}{m_{\wt{e}R}^2} \sum_a \kappa_3 |\lambda_3^a|^2 M_a \sim {\cal O}(1) \,.
\eeq
This is achieved when $v_R \sim m_{\wt e R} \sim M_a$ and $\kappa_3|\lambda^a_3|^2\sim1$, as shown in Fig.~\ref{fig:charged}.
The muon and electron yukawa couplings are related to the tau yukawa coupling as 
\beq
\frac{(y_e)_i}{y_\tau} \simeq \frac{\sum_a \kappa_i |\lambda_i^a|^2 M_a}{\sum_a \kappa_3 |\lambda_3^a|^2 M_a} \,,
\eeq
up to the logarithmic corrections. 
If all couplings and masses have comparable size, 
$\frac{(y_e)_i}{y_\tau} \simeq {\cal O}(1)$ is predicted, but it is not realistic. 
To achieve the realistic charged lepton spectrum, we need some non-trivial structure for $M_a$ and $\lambda_i^a$.

In this work, we consider the case where $M_a$ is hierarchical (say $M_1 \ll M_2 \ll M_3$). 
Since the heavier $n^a$ have larger contribution in this case, 
the mass of the charged lepton $e_i$ is generated mostly from the $a=3$ diagram unless there is a significant hierarchy between $\lambda^3_i$ and the other couplings $\lambda^{1,2}_i$. 
Then, we obtain $\frac{(y_e)_i}{y_\tau} \sim \frac{\kappa_i|\lambda_i^3|^2}{\kappa_3|\lambda_3^3|^2}$. 
This indicates an upper limit on the $n^3$ contribution to the muon and electron masses: $|\lambda^3_i| \lesssim \sqrt{\frac{(m_e)_i}{m_\tau}}$ for $\kappa_i\sim1$.
Similarly, we have an upper limit on the $n^2$ contribution to the electron mass: $|\lambda^2_1| \lesssim \sqrt{\frac{M_3}{M_2}\frac{m_e}{m_\mu}}$ for $\kappa_i\sim1$.
In contrast, if $\lambda^a_i$ for $i<a$ are very suppressed compared with these upper limits, 
the mass of the charged lepton $e_i$ is generated mostly from the $i=a$ diagram and we can roughly determine the ratio of $M_a$. 
For illustration, let us consider the following coupling structure,
\beq
\left(\lambda^a_i\right) = 
\begin{pmatrix} \lambda^1_1 & \lambda^1_2 & \lambda^1_3 \\ 
0 & \lambda^2_2 & \lambda^2_3 \\ 
0 & 0 & \lambda^3_3 
\end{pmatrix}.
\label{eq:Yukawa-lambda}
\eeq
Then, taking $\kappa_i|\lambda_i^a|^2=1$ for all non-vanishing components, the mass spectrum of $n^a$ is related directly to the ratio of the charged lepton masses: 
\beq
\frac{M_i}{M_3} \sim \frac{(m_e)_i}{m_\tau} 
\sim \left\{ \begin{array}{ll} 0.1 &~~~\mbox{for $i=2$} \\ 0.0005 &~~~\mbox{for $i=1$} \end{array} \right. \,.
\eeq
Given that a smaller $\kappa_i|\lambda_i^a|^2$ leads to a heavier $M_a$, 
we have a rough lower bound, $M_i \gtrsim \frac{(m_e)_i}{m_\tau} M_3$. 
Combined it with $M_3 \sim v_R \gtrsim 20\,\TeV$ (see Sec.~\ref{sec:Neff} for the detail of this lower limit), the lightest fermion $n^1$ can be as light as GeV.
Hereafter, we assume the coupling structure like Eq.~(\ref{eq:Yukawa-lambda}), so that the mass spectrum of $n^a$ is determined through the charged lepton masses.

It is easy to confirm our approximate prediction in Fig.~\ref{fig:charged}. 
Namely, $(y_e)^{a}_{i}$ is proportional to $M_a$ for $M_a\ll m_{\wt e R}$, up to the minor logarithmic corrections. 
In this region, $(y_e)^{a}_{i}$ are naturally small, and we can use Eq.~(\ref{eq:ye_approx}) to fit the muon and electron yukawa couplings. 
If we take $\kappa_{i} |\lambda^{a}_{i}|^2=1/x$ instead of $\kappa_{i} |\lambda^{a}_{i}|^2=1$, the $M_a$ value required to fit the charged lepton mass is shifted approximately as $M_a=xM_{a,0}$, where $M_{a,0}$ denotes the mass of $n^a$ for $\kappa_{i} |\lambda^{a}_{i}|^2=1$.
Our approximation in Eq.~(\ref{eq:ye_approx}) is not valid for the tau yukawa coupling, since $M_a$ are not sufficiently small in this case. 
We need to use the full loop function in Eq.~(\ref{eq:loop_function}) to evaluate it. 
Note that there is a constraint to avoid the tachyonic mass of charged scalars, 
\beq
m^2_{\widetilde e L} m^2_{\widetilde e R} > \kappa_i^2 v^2_L v^2_R \,.
\eeq
In our analysis, $m_{\wt e L}$ is assumed to be ${\cal O}(v_L)$, which means $m_{\widetilde e R}$ has to be ${\cal O}(v_R)$ or heavier, because of this constraint.
For the same reason, $\kappa_i$ cannot be arbitrarily large.

\subsection{Neutrino mass matrix}
\label{sec:Mnu_fit}

From various sets of neutrino oscillation data, 
the neutrino mass squared differences and neutrino mixing angles are measured with good accuracy. 
Assuming the normal ordering (NO) and no charged lepton mixing, 
we can express the neutrino mass matrix in terms of the PMNS matrix elements $U_{ij}$ and the neutrino masses $m_i$. 
When $m_1$ is vanishing, we find it takes the form, 
\begin{equation}
(\Hat M^{\rm obs}_\nu)_{ij} = 0.031\,\eV\begin{pmatrix}
0.12  & -0.055+0.094\,i & -0.22+0.084\,i  \\  
-0.055-0.094\,i & 1  & 0.68-0.011\,i \\  
-0.22-0.084\,i  & 0.68+0.011\,i & 0.81 \end{pmatrix} ,
\label{eq:neutrino}
\end{equation}
where the central values in Table \ref{tab:parameter} are used. 
For a finite $m_1$, the hierarchy among the matrix elements becomes smaller, 
so that a small $m_1$ may be favored in our model.
In the following, 
we try to reproduce this NO mass matrix with tiny $m_1$ in our model.
In Fig.\,\ref{fig:model1} - Fig.\,\ref{fig:model3_2},
the corresponding values of $\Hat M^{\rm obs}_\nu$ are depicted by black bands. The bands are drawn, varying $m_1$ from $m_1=0$ to $m_1=0.01$\,eV and taking into account the $1 \sigma$ errors in Table \ref{tab:parameter}. 
In the case of the inverted ordering (IO), 
there is a larger gap between the neutrino and charged lepton mass matrix structures, compared with the NO case, and it is harder to fill that gap in our model. 
Thus, we do not consider the IO case in this work. 

\begin{table}[t]
\begin{center}
\begin{tabular}{|cc|cc|}
\hline
Observables & Values  & Observables & Values \\ \hline
$m_e$ & 0.510 $\times 10^{-3}$ GeV & $m_\mu$ & 0.1057 GeV \\
$m_\tau$ & 1.777 GeV  &&  \\ \hline
$\Delta m^2_{21}$ & $(7.53 \pm 0.18) \times 10^{-5}$ eV$^2$  & $\sin^2 \theta_{12} $ & $0.307 \pm 0.013$ \\
$\Delta m^2_{32}$ (NO) & $(2.455 \pm 0.028) \times 10^{-3}$ eV$^2$  & $\sin^2 \theta_{23} $ (NO)& $0.558^{+0.015}_{-0.021}$ \\
$\Delta m^2_{32}$ (IO) & $(-2.529 \pm 0.029) \times 10^{-3}$ eV$^2$ & $\sin^2 \theta_{23} $ (IO)& $0.553^{+0.016}_{-0.024}$ \\
$\delta/\pi$  &  $1.19 \pm 0.22 $ & $\sin^2 \theta_{13} $ & $(2.19 \pm 0.07) \times 10^{-2}$ \\ \hline
\end{tabular}
\end{center}
\caption{The values of the lepton observables we use. 
These values are all taken from Ref.~\cite{ParticleDataGroup:2024cfk}.}
\label{tab:parameter}
\end{table}

Let us have a close look at the neutrino mass matrix elements. 
For $m^2_{S_L\alpha} \ll m^2_{S_R\beta}$,
we can approximate $(\hat M_\nu)_{ij}$ to 
\beq
(\Hat M_\nu)_{ij} \simeq v_L v_R \left\{ \sum^{3}_{a=1} \sum_{\alpha,\beta} \sum_{S_L,S_R} \frac{(\lambda_{S_L})^{a}_{i\alpha} \left(\kappa_\nu^{S_L S_R}\right)_{\alpha\beta}(\lambda_{S_R}^{*})^{a}_{j\beta} }{4\,(4 \pi)^2} 
\frac{M_a}{m^2_{S_R\beta}} \log\frac{M_a^2}{m^2_{S_R\beta}}\right\}.
\label{eq:Mnu_approx}
\eeq
Here, $\kappa_\nu^{\rm equiv}$ in $\kappa_\nu^{S_LS_R}=V_{S_L}\kappa_\nu^{\rm equiv} V_{S_R}^\dagger$ is induced from the two-loop diagrams (A) and (B). 
Under our assumptions (\ref{eq:assumption1})-(\ref{eq:assumption2}), $\kappa_\nu^{\rm equiv}$ is flavor diagonal, i.e. $(\kappa_\nu^{\rm equiv})_{\alpha\beta}\propto\kappa_\alpha\,\delta_{\alpha\beta}$.
See Appendix \ref{appendix} for the details. 
To realize the small neutrino mass scale of ${\cal O}(0.01\,\eV)$, 
$\kappa_\nu^{\rm equiv}$ is required to be smaller than ${\cal O} (10^{-10})$. 
That small $\kappa_\nu^{\rm equiv}$ can be achieved in our model by adjusting the value of $m^U_{33}/m_T$, which can be chosen independently of the top quark mass.

Given the coupling structure in Eq.\,(\ref{eq:Yukawa-lambda}), 
$(\Hat M_\nu)_{1i}$ ($i=1,2,3$) is generated solely from the $a=1$ diagram, while $(\Hat M_\nu)_{22,23}$ and $(\Hat M_\nu)_{33}$ are respectively generated from the $a=2$ and $a=3$ diagrams. 
As a result, our model predicts 
$(\Hat M_\nu)_{1i}\propto M_1$, 
$(\Hat M_\nu)_{22,23}\propto M_2$ and $(\Hat M_\nu)_{33}\propto M_3$. 
Since $M_1$ should be much smaller than $M_{2,3}$ to reproduce the electron mass, 
the neutrino mass matrix tends to be hierarchical, $|(\Hat M_\nu)_{1i}| \ll |(\Hat M_\nu)_{22,23,33}|$.
To diminish this hierarchy, we introduce a large mass hierarchy to $S_R^\beta$. 
If the lightest $S_R^\beta$ (which we take $S_R^1$) contributes dominantly to $|(\Hat M_\nu)_{1i}|$ but not to $|(\Hat M_\nu)_{22,23,33}|$, the former receives enhancement by a factor of $\frac{m_{S_R2}^2}{m_{S_R1}^2}$ or $\frac{m_{S_R3}^2}{m_{S_R1}^2}$, compared with the latter two.
If this enhancement compensates the suppression from the $M_a$ hierarchy, 
we will be able to obtain the realistic neutrino mass matrix structure. 
In Eq.~(\ref{eq:Mnu_approx}), we assumed $m^2_{S_L\alpha} \ll m^2_{S_R\beta}$, 
but this compensation can also happen in the opposite case, $m^2_{S_R\alpha} \ll m^2_{S_L\beta}$.

Below, we demonstrate that the realistic neutrino mass matrix structure can be achieved with the mass hierarchy of the neutral scalars. 
We consider two benchmark scenarios, 
\begin{itemize}
\item 
light $h_R$ scenario with $m_{h_R1} \ll m_{h_L1} \sim {\cal O}(v_L)$ ,
\item 
light $h_L$ scenario with $m_{h_L1} \ll m_{h_R1} \sim {\cal O} (v_L)$.
\end{itemize}
In the former case, $ m_{h_R1} = m_{A_R1}$ is also assumed in our analysis. 
In the latter case, on the other hand, we have to make $m_{h_L1} \ll m_{A_L1}$ to avoid the EW precision bound.

\subsubsection{Light $h_R$ scenario}
\label{sec:light_hR}

First, we consider the case where $h_R^1$ and $A_R^1$ are lighter than the other scalars. 
These two light scalars are assumed to be degenerate,
\beq
m_{h_R1} = m_{A_R1} =: m_{\wt\nu_R1}\,.
\eeq
We further simplify the analysis by considering the degenerate spectrum for the other extra scalars: 
\beq
m_{h_Li} = m_{A_Li} =: m_{\wt\nu L}\,,\quad
m_{\wt\nu R2} = m_{\wt\nu R3} \,,
\eeq
where $m_{h_RI} = m_{A_RI}=: m_{\wt\nu RI}$ ($I=2, \,3 $).
The spectrum of our interest in this scenario is as follows:
\beq
m_{\wt\nu L} \sim m_{\wt eL} \sim {\cal O}(v_L) \,,\quad
m_{\wt\nu R1} \ll m_{\wt\nu R2} \sim m_{\wt eR} \sim {\cal O}(v_R) \,. 
\eeq

Let us look at the neutrino mass matrix elements.
For illustration, we take $(\kappa_\nu^{\rm equiv})_{\alpha\beta}=\left(1.0 \times 10^{-11}\right)\delta_{\alpha\beta}$ and $v_R=25\,\TeV$, and choose several mass parameters to be 
\beq
m_{\wt eL}=500\,\GeV,\quad
m_{\wt eR}=12.5\,\TeV,\quad
M_2=95\,\GeV,\quad
M_3=4.25\,\TeV\,.
\label{eq:inputs_lighthR}
\eeq
These masses are chosen to reproduce the muon and tau masses with $\kappa_2|\lambda^2_2|^2=\kappa_3|\lambda^3_3|^2=1$. 
All non-vanishing couplings $\lambda^a_{i}$ except $|\lambda^1_{1}|$ are fixed to unity.
In our analysis, instead of giving the concrete scalar mass matrices, we simply parameterize the scalar mixing matrices as  
\beq
V_{h_L} = V_{A_L} = \1 \,,\quad
V_{h_R} = V_{A_R} 
	= \begin{pmatrix} 
		\cos\theta & \sin\theta & 0 \\ 
		- \frac{1}{\sqrt2} \sin\theta & \frac{1}{\sqrt2} \cos\theta & \frac{1}{\sqrt2} \\
		\frac{1}{\sqrt2} \sin\theta & - \frac{1}{\sqrt2} \cos\theta & \frac{1}{\sqrt2} 
	\end{pmatrix} .
\label{eq:mixing_model1}
\eeq
The $(i,1)$ matrix elements $\left(V_{S_R}\right)_{i1}$ contributes dominantly to $(\Hat M_\nu)_{1i}$. 

\begin{figure}[t]
\centering
\includegraphics[width=7cm]{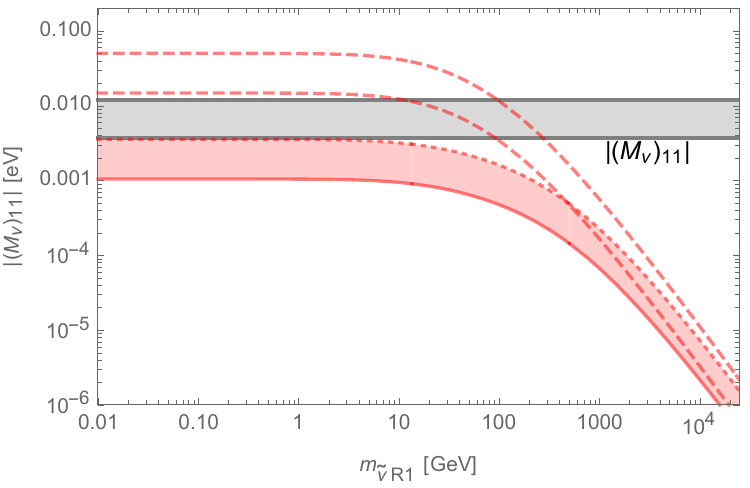}\includegraphics[width=7cm]{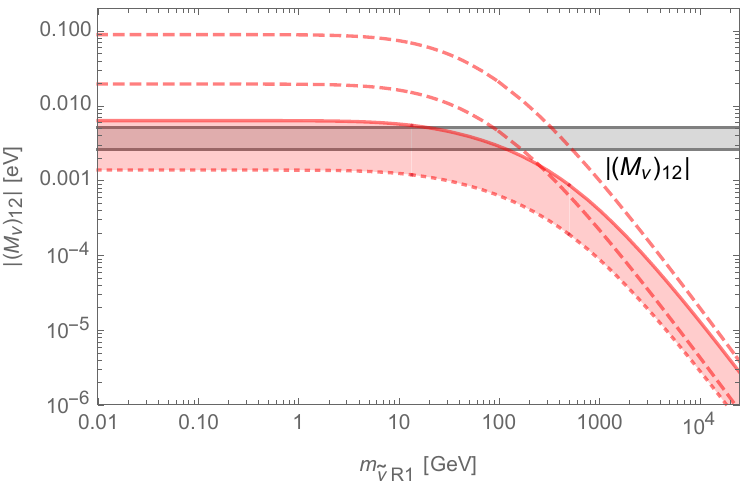}
\includegraphics[width=7cm]{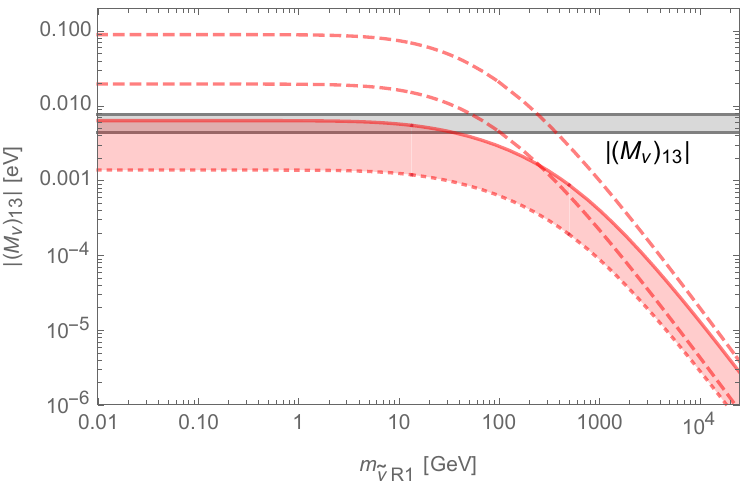}
\caption{The model predictions for the neutrino mass matrix elements $|(\hat M_\nu)_{1j}|$ in the light $h_R$ scenario.
The red bands show the predictions with $m_{\wt\nu L}=500$\,GeV and $\theta\in[0.1,\,1]$, where the solid (dotted) red lines correspond to $\theta=1$ (0.1). 
For $m_{\wt\nu L}=100$\,GeV, the red bands shift upward, as indicated by two red dashed lines. The horizontal black bands show the observed values within 1$\sigma$ ($0 \leq m_1 \leq 0.01$\,eV).
}\label{fig:model1}
\end{figure}

\begin{figure}[t] 
\centering
\includegraphics[width=7cm]{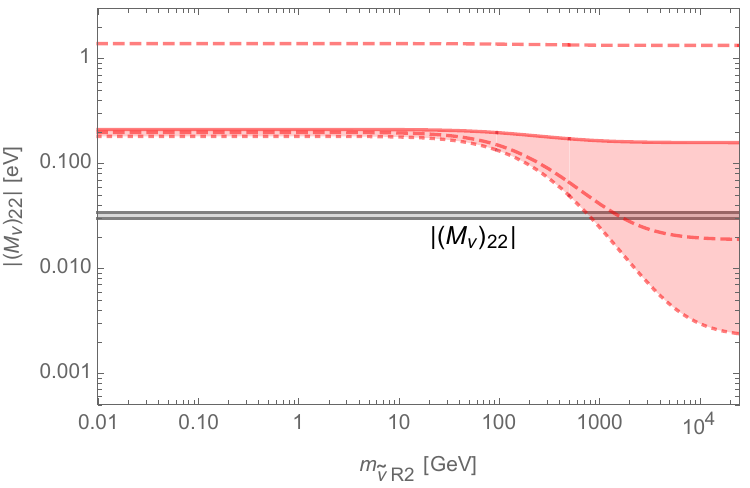}\includegraphics[width=7cm]{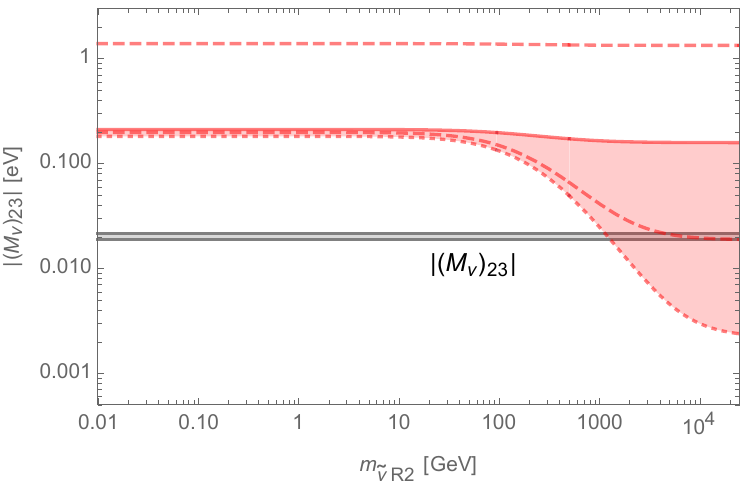}
\includegraphics[width=7cm]{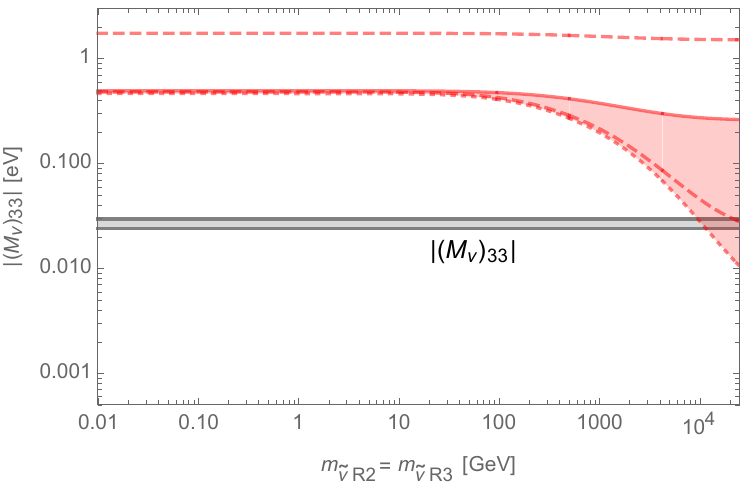}
\caption{The model predictions for the neutrino mass matrix elements $|(\hat M_\nu)_{22}|$, $|(\hat M_\nu)_{23}|$ and $|(\hat M_\nu)_{33}|$ in the light $h_R$ scenario. 
The red bands show the predictions with $m_{\wt\nu_L}=500$\,GeV and $\theta\in [0.1,\,1]$, where the solid (dotted) lines correspond to $\theta=1$ (0.1).
For $m_{\wt\nu_L}=100$\,GeV, the red bands shift upward, as indicated by two red dashed lines. 
The horizontal black bands show the observed values within 1$\sigma$ ($0 \leq m_1 \leq 0.01$\,eV).}
\label{fig:model1_2}
\end{figure}

In Fig.~\ref{fig:model1}, 
we show the model predictions for the neutrino mass matrix elements $|(\hat M_\nu)_{i1}|$ with $M_1=13.5\,\GeV$.
$\kappa_1|\lambda^1_1|^2$ is fixed to reproduce the electron mass.
In the figure, we take into account the contribution from $\wt\nu_R^1$ but ignore the ones from $\wt\nu_R^2$ and $\wt\nu_R^3$, since the latter ones are suppressed by their masses.
The red bands show the predictions with $m_{\wt\nu L}=500$\,GeV and $\theta\in[0.1,\,1]$, where the solid (dotted) lines correspond to $\theta=1$ (0.1). 
For $\theta<0.1$, $|(\Hat M_\nu)_{11}|$ is not sensitive to $\theta$, 
while $|(\Hat M_\nu)_{12}|$ and $|(\Hat M_\nu)_{13}|$ become smaller as $\theta$ gets smaller. 
In contrast, for $\theta>1$, $|(\Hat M_\nu)_{11}|$ is more suppressed as $\theta$ gets larger. 
$|(\Hat M_\nu)_{12}|$ and $|(\Hat M_\nu)_{13}|$ are maximized around $\theta=1$.
It is observed that $|(\Hat M_\nu)_{i1}|$ strongly depend on $m_{\wt\nu R1}$ and approach the observed values for $m_{\wt\nu R1} \lesssim 100$ GeV. 
$|(\Hat M_\nu)_{i1}|$ is also sensitive to $m_{\wt\nu L}$. 
If we take $m_{\wt\nu L}=100$\,GeV, the red regions shift upward, as indicated by two red dashed lines. 

The other matrix elements depend mainly on $m_{\wt\nu R2}$.
In Fig.~\ref{fig:model1_2}, we show the model predictions for $|(\Hat M_\nu)_{22}|$, $|(\Hat M_\nu)_{23}|$, and $|(\Hat M_\nu)_{33}|$ with $m_{\wt\nu R1}=M_1=13.5\,\GeV$.
It is easy to see that these matrix elements are consistent with the observed values only if $m_{\wt\nu R2}$ are large.

\subsubsection{Light $h_L$ scenario}
\label{sec:light_hL}

\begin{figure}[t]
\centering
\includegraphics[width=7cm]{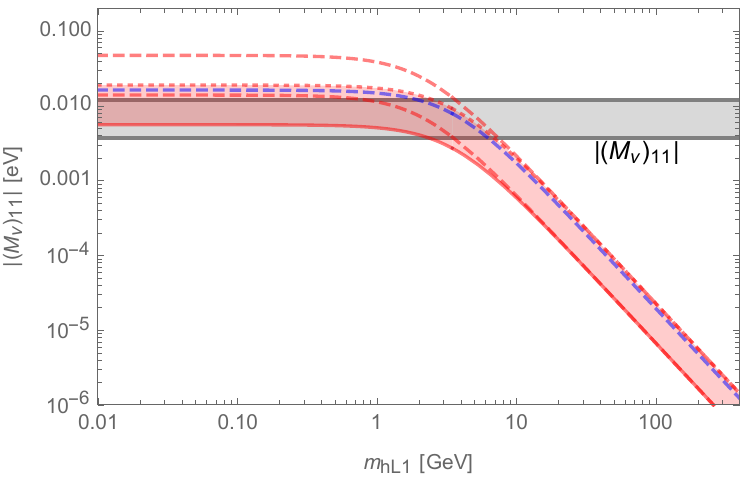}\includegraphics[width=7cm]{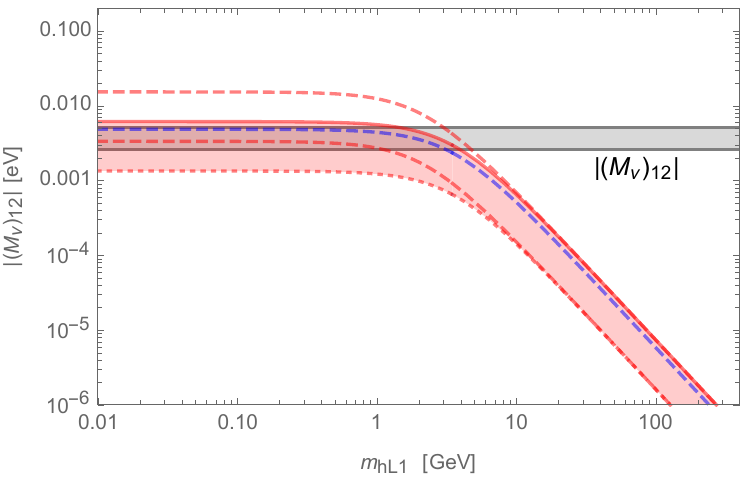}
\includegraphics[width=7cm]{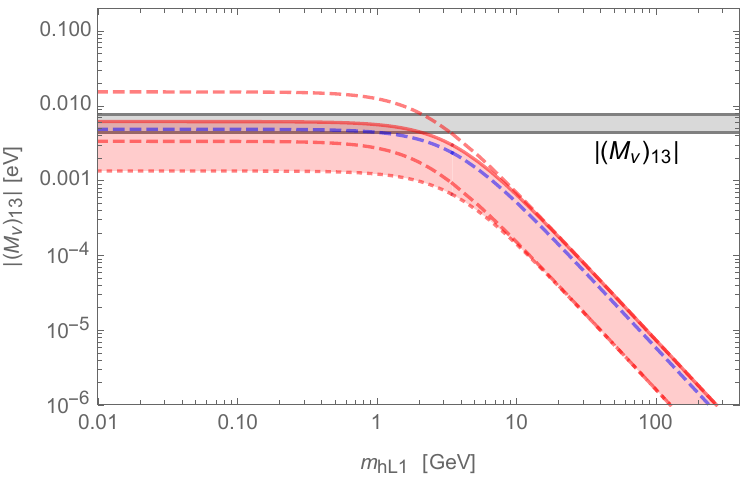}
\caption{The model predictions for the neutrino mass matrix elements $|(\hat M_\nu)_{1j}|$ in the light $h_L$ scenario.
The red bands show the predictions with $\sqrt{\Delta m_R^2}=500$\,GeV and $\theta\in[0.1,\,1]$, where the red solid (dotted) lines correspond to $\theta=1$ (0.1). 
When we take $\sqrt{\Delta m_R^2}=300$\,GeV, the red bands shift upward, as indicated by two red dashed lines. 
The black bands in the plots show the observed values within 1$\sigma$ ($0 \leq m_1 \leq 0.01$\,eV).
}\label{fig:model3}
\end{figure}

\begin{figure}[t]
\centering
\includegraphics[width=7cm]{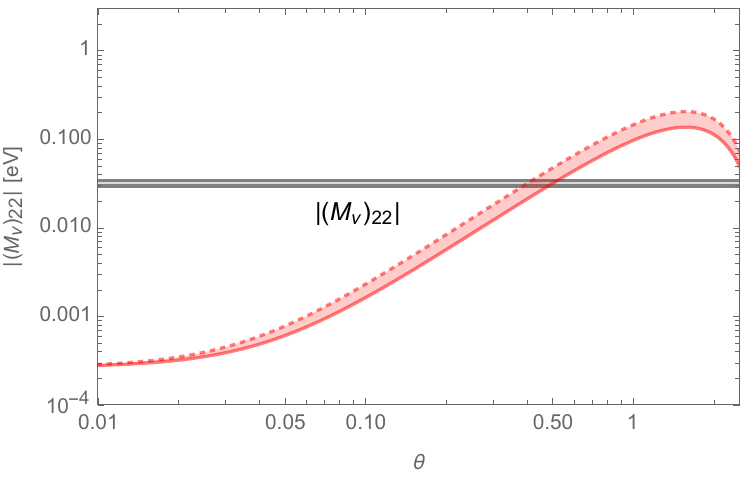}\includegraphics[width=7cm]{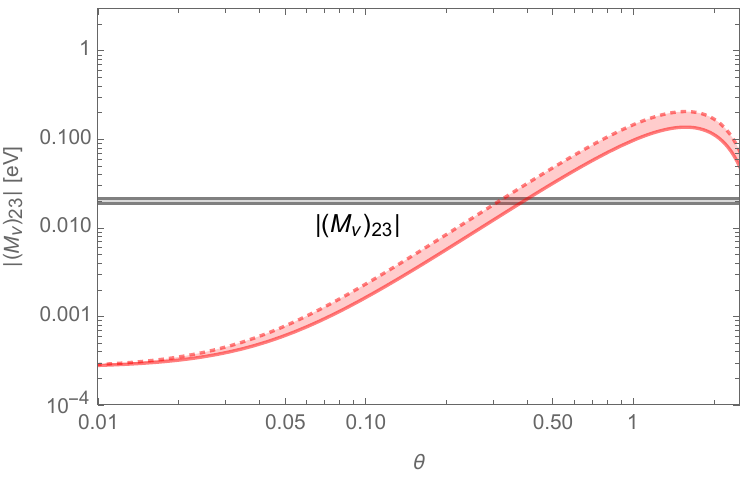}
\includegraphics[width=7cm]{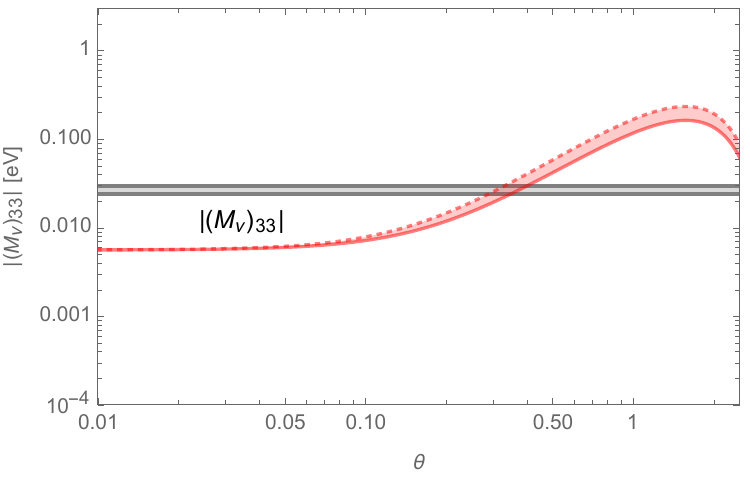}
\caption{The model predictions for the neutrino mass matrix elements, $|(\hat M_\nu)_{22}|$, $|(\hat M_\nu)_{23}|$ and $|(\hat M_\nu)_{33}|$ in the light $h_L$ scenario. 
The red bands show the predictions with $\sqrt{\Delta m_R^2} \in [300,\,500]\GeV$, where the solid (dotted) line corresponds to $\sqrt{\Delta m_R^2}=300$ (500) GeV.
The horizontal black bands show the observed values within 1$\sigma$ ($0 \leq m_1 \leq 0.01$\,eV).
}\label{fig:model3_2}
\end{figure}

Next, we consider the case where $h_L^1$ is lighter than the other scalars. 
As we will see below, $h_L^1$ should be as light as 10\,GeV to explain the neutrino oscillation data. 
If $A_L^1$ is also light in this case, the model is faced with the LEP bound \cite{Cao:2007rm, Lundstrom:2008ai}. 
Thus we take a very peculiar spectrum where only $h_L^1$ is light and $A_L^1$ has the EW scale mass. 
It is shown in Refs.~\cite{Okawa:2020jea, Iguro:2022tmr} that such a spectrum is still viable. 

In our analysis, we consider the scalar mixing matrices in the form, 
\beq
V_{h_L} = V_{h_R}  =\begin{pmatrix}
 \cos \theta & \sin \theta& 0   \\
- \frac{1}{\sqrt{2}} \sin \theta & \frac{1}{\sqrt{2}} \cos \theta& \frac{1}{\sqrt{2}} \\
 \frac{1}{\sqrt{2}} \sin \theta & - \frac{1}{\sqrt{2}} \cos \theta & \frac{1}{\sqrt{2}}
\end{pmatrix}
\label{eq;model3_mixing},\quad
V_{A_L} = V_{A_R} = \1 \,.
\eeq
For illustration we take $(\kappa_\nu^{\rm equiv})_{ij}= \left(2.0\times10^{-11}\right) \delta_{ij}$ and $v_R=25\,\TeV$, and choose several mass parameters to be 
\beq
m_{\wt eL}=300\,\GeV,\quad
m_{\wt eR}=43\,\TeV,\quad
M_2=975\,\GeV,\quad
M_3=43\,\TeV.
\label{eq:inputs_lighthL}
\eeq
These masses are consistent with the observed muon and tau masses, but 
a larger $m_{\wt eR}$ forces $\lambda^3_3$ to be larger: $\lambda^3_3=\sqrt{1.67}$. 
All other non-vanishing couplings $\lambda^i_a$ are fixed to unity. 
Regarding the neutral scalar spectrum, we assume for simplicity 
\beq
m_{h_L1} \ll m_{A_L1} = m_{h_L2} = m_{A_L2} = m_{h_L3} = m_{A_L3} =: m_{\wt\nu L23} \,,
\eeq
and 
\beq
m_{h_R1} \ll m_{A_R1} = m_{h_R2} = m_{A_R2} = m_{h_R3} = m_{A_R3} =: m_{\wt\nu R23} \,.
\eeq
The mass spectrum of our interest is thus
\beq
m_{h_L1} \ll m_{\wt\nu L23} \sim m_{\wt eL} \sim {\cal O}(v_L) \,,\quad
m_{\wt\nu R1} \sim {\cal O}(v_L) \ll m_{\wt\nu R23} \sim m_{\wt eR} \sim {\cal O}(v_R) \,.
\eeq
Note that to realize this mass spectrum, we need to introduce a small mass splitting between $\wt e_R^1$ and $\wt e_R^{2,3}$. 
This is understood from the relation among the masses of $h_L^1$, $h_R^1$, $\wt e_L$, and $\wt e_R$, 
\begin{equation}
\frac{m^2_{\wt eR}-m^2_{h_R1}}{v_R^2}+\frac{\Delta m^2_R}{v_R^2}=\frac{m^2_{\wt eL}-m^2_{h_L1}}{v^2_L} \,.
\end{equation}
where $\Delta m^2_R=m_{\wt e_R1}^2-m_{\wt e_R}^2$ denotes the squared mass difference between $\wt e_R^1$ and $\wt e_R^{2,3}$. 
When $h_L^1$ is much lighter than the EW scale, $h_R^1$ also need to be light if there is no mass splitting $\Delta m_R^2$.
In particular, $m_{h_R1}$ becomes negative with our mass inputs Eq.~(\ref{eq:inputs_lighthL}) and $m_{h_L1}\ll v_L$. 
To keep $h_R^1\sim{\cal O}(v_L)$, therefore, we introduce this small mass splitting of ${\cal O}(v_L)$. 

In Fig.~\ref{fig:model3}, 
we show the model predictions for the neutrino mass matrix elements $|(\hat M_\nu)_{1j}|$ with $M_1=3.48\,\GeV$ and $m_{\wt\nu L23}=300$\,GeV. 
The red bands show the predictions with $\sqrt{\Delta m_R^2} =500$\,GeV and $\theta\in[0.1,\,1]$, where the red solid (dotted) lines correspond to $\theta=1$ (0.1). 
We highlight with the blue dashed line our predictions in the case of $\theta=0.4$, which provides a good fitting to $|(\hat M_\nu)_{22}|$, $|(\hat M_\nu)_{23}|$ and $|(\hat M_\nu)_{33}|$ (see Fig.~\ref{fig:model3_2}). 
In the plots, we only take into account the contribution from $h_{L,R}^1$ and ignore the ones from $h_{L,R}^{2,3}$, since the latter ones are subdominant due to their heaviness. 
It is seen that the model can explain the observed values for $m_{h_L1} \lesssim 10$\,GeV. 
For $\sqrt{\Delta m_R^2}=300$\,GeV, the red bands shift upward, as indicated by two red dashed lines. 

In Fig.~\ref{fig:model3_2}, we show the model predictions of $|(\hat M_\nu)_{22}|$, $|(\hat M_\nu)_{23}|$ and $|(\hat M_\nu)_{33}|$ as a function of $\theta$. 
We take $m_{h_L1}=M_1=3.48\,\GeV$ in the figure. 
The red bands show the predictions with $\sqrt{\Delta m_R^2} \in [300,\,500]\GeV$ where the solid (dotted) line corresponds to $\sqrt{\Delta m_R^2}=500$ ($300$) GeV.
With this parameter set, our model can fit the experimental results when $\theta\simeq0.4$.

\section{Cosmological bound from right-handed neutrinos}
\label{sec:Neff}

Since the neutrinos are Dirac fermions in our model, 
their right-handed components can contribute to the radiation density in the universe if they are thermalized with the SM plasma at high temperatures. 
The extra contribution to the radiation density, also known as dark radiation, modifies the expansion history of the universe and alters the predictions for the primordial nucleosynthesis and the cosmic microwave background anisotropies. 
In this section, we estimate the right-handed neutrino contribution to the radiation density and evaluate the resultant cosmological bound in our model. 

The radiation density of the universe in the presence of dark radiation is given by 
\begin{align}
\rho_{\rm rad} 
	& = \rho_\gamma + \rho_\nu + \rho_{\rm DR}^{} 
	= \frac{\pi^2}{30} T_\gamma^4 \left\{ g_\gamma + \frac{7}{8} g_\nu N_\eff \(\frac{T_\nu}{T_\gamma}\)^{4/3} \right\} \,,	
\label{eq:rad_den}
\end{align}
where $g_\gamma=2$ and $g_\nu=2$ denote the spin degrees of freedom for photon and left-handed neutrinos, respectively. 
The energy density of dark radiation is rendered in a shift of $N_\eff$, and 
for a particle species $X$ constituting dark radiation, 
its contribution to $\Delta N_\eff$ is expressed as 
\beq
\Delta N_\eff 
	:=  N_\eff - N_{\eff,\nu} 
	= \sum_X \frac{8}{7} \frac{g_X^{}}{g_\nu} \(\frac{T_X}{T_\nu}\)^{4/3} 
	= \sum_X \frac{4}{7} g_X^{} \(\frac{11}{4}\)^{4/3} \(\frac{T_X}{T_\gamma}\)^{4/3} \,,
\eeq
where $g_X^{}$ denotes the spin degrees of freedom (times 7/8 for fermions) and $T_\nu = (4/11)^{1/3} T_\gamma$ is used in the last equality. 
If $X$ is in thermal equilibrium with the SM plasma at some high temperatures, 
the $X$-to-photon temperature ratio is determined from the entropy conservation, given in terms of the ratio of the effective entropy degrees of freedom, 
\beq
\frac{T_X}{T_\gamma} = \(\frac{g_{*s}(T_0)}{g_{*s}(T_{D,X})}\)^{1/3} \,,
\eeq
where $g_{*,s}(T_0) = 43/11$ and $g_{*,s}(T_{D,X})$ denotes the sum of all relativistic degrees of freedom except $X$ in the thermal plasma at the time of its thermal decoupling. 
Putting all known numbers, we end up with 
\beq
\Delta N_\eff \simeq \sum_X  0.0268 \, g_X^{} \(\frac{106.75}{g_{*s}(T_{D,X})}\)^{4/3} \,.
\label{eq:DeltaNeff}
\eeq

\begin{figure}[t]
\centering
\includegraphics[width=0.5\textwidth]{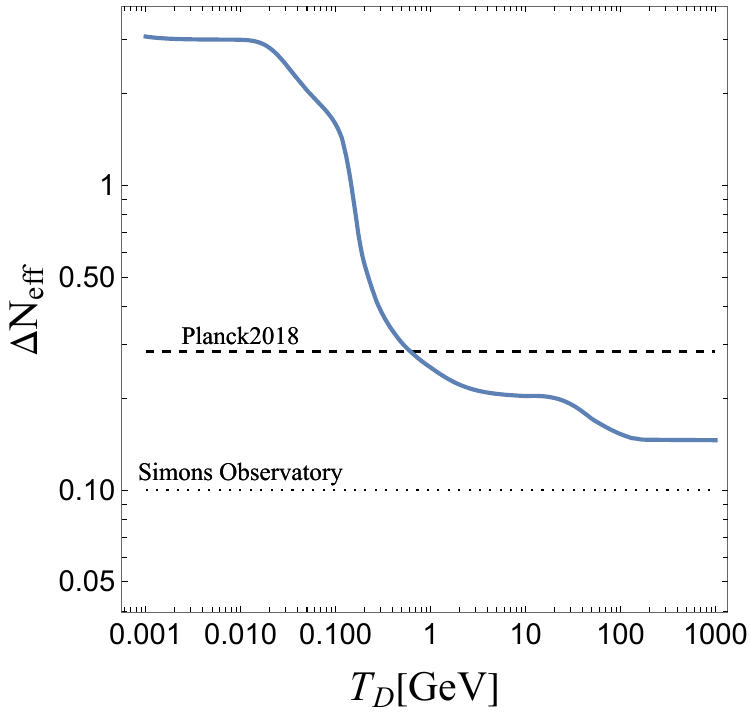}
\caption{The contribution to $\Delta N_\eff$ from three right-handed neutrinos.
The black dashed and dotted lines correspond respectively to the current Planck+BAO bound \cite{Planck:2018vyg} and the projected $2\sigma$ limit of Simons Observatory \cite{SimonsObservatory:2018koc,SimonsObservatory:2019qwx}.}
\label{fig:Neff}
\end{figure}

The contribution from three right-handed neutrinos is calculated from Eq.~(\ref{eq:DeltaNeff}) by taking $g_X^{} = \frac{7}{4} N_{\nu_R}$ (with $N_{\nu_R}=3$). 
Figure \ref{fig:Neff} shows $\Delta N_\eff$ from three right-handed neutrinos as a function of the decoupling temperature $T_D$, assuming they are decoupled at the same time. 
The black dashed and dotted lines correspond respectively to the current Planck+BAO bound \cite{Planck:2018vyg} and the projected $2\sigma$ limit of Simons Observatory, $\Delta N_{\rm{eff}}<0.1$ \cite{SimonsObservatory:2018koc,SimonsObservatory:2019qwx}. 
The Planck bound excludes $T_D \lesssim 600\,\MeV$, 
see also \cite{Babu:2022ikf, Borah:2025fkd}. 

The decoupling temperature of a right-handed neutrino $\nu_R$ is approximately evaluated using the Gamov criterion, 
\beq
\Gamma_{\nu_R}(T_D) \simeq H(T_D) \,,
\eeq
where $H=\sqrt{\frac{8\pi^3}{90} g_*(T)} \frac{T^2}{M_{\rm pl}}$ is the Hubble rate, and the reaction rate is given by 
\beq
\Gamma_{\nu_R} \equiv \sum_{j,k,l} \VEV{\sigma v_r}_{\nu_R j \to kl} n^{\rm eq}_j \,.
\eeq
Here, $\vev{\sigma v_r}_{X \to Y}$ denotes the thermal averaged cross section, 
\beq
\vev{\sigma v_r}_{ij \to kl} \equiv \frac{T}{32\pi^4} \frac{1}{n_i^{\rm eq}(T)\,n_j^{\rm eq}(T)} \int_{s_{\rm min}}^\infty \mathrm{d}s \, \sigma_{ij \to kl} \, \frac{\lam(s,M_i^2,M_j^2)}{\sqrt{s}} \, K_1(\sqrt{s}/T) \,,
\label{eq:sigmavT}
\eeq
where $s_{\rm min}={\rm Max}\,[(M_i+M_j)^2, (M_k+M_l)^2]$ and $\sigma_{ij \to kl}$ is the cross section for the $ij \to kl$ process. 
In our model, the thermalization can occur mainly through the $W_R$ and $Z'$ exchange processes. 
The relevant thermalization processes are 
\begin{itemize}
\item annihilation: $\nu_R^i \bar{\nu}_R^i \to f \bar{f}$ ($f\neq \nu_R$)
\item scattering: $\nu_R^i f \to \nu_R^i f$, $\nu_R^i \bar{f} \to \nu_R^i \bar{f}$ ($f\neq \nu_R$)
\item co-annihilation: $\nu_R^i \bar{e}^i \to u^j \bar{d}^j$ 
\item conversion: $\nu_R^i \bar{e}^i \to \nu_R^j \bar{e}^j$ ($i\neq j$).
\end{itemize}
Taking for simplicity the massless limit of the fermions involved in the scattering, we obtain the scattering rate, 
\beq
\VEV{\sigma v_r}_{X\to Y}
	& \simeq C_{X\to Y} \, \frac{8 G_F^2 T^2}{\pi} \(\frac{v_L}{v_R}\)^4 \,, 
\eeq
where $C_{X\to Y}$ is given by
\beq
C_{\nu_R^i \bar{\nu}_R^i \to f \bar{f}} & = 2 N_{c,f} \left\{ (\delta_{fe}-g'_{R})^2 + (g'_L)^2 \right\} \quad (f\neq\nu_R) \,,\\
C_{\nu_R^i f \to \nu_R^i f} & = \left\{ 3 (\delta_{fe}-g'_{R})^2 + (g'_L)^2 \right\} \,,\\
C_{\nu_R^i \bar{f} \to \nu_R^i \bar{f}} & = \left\{ (\delta_{fe}-g'_{R})^2 + 3 (g'_L)^2 \right\} \,,\\
C_{\nu_R^i \bar{e}^i \to u^j \bar{d}^j} & = N_c \, C_{\nu_R^i \bar{e}^i \to \nu_R^j \bar{e}^j} = N_c \,.
\eeq
Here, $N_c=3$ and $N_{c,f}$ denotes the color of a fermion species $f$, and $g'_L$ and $g'_R$ are defined by 
\beq
g'_L=-s_R^2 Y_{f_L}, \quad
g'_R=T^3_{R,f_R}-s_R^2 Y_{f_R} \,,
\eeq
with $Y_{f_L}$ $(Y_{f_R})$ the hypercharge of $f_L$ $(f_R)$, 
$s_R^{}=2g_{B-L}^{}/\sqrt{g_R^2+4g_{B-L}^2}$, and $(g_Y^{})^{-2}=(g_R^{})^{-2}+(2g_{B-L}^{})^{-2}$. 
From the left-right symmetry, we have $g:=g_L^{}=g_R^{}$, which leads to  
\beq
s_R^2 = \frac{4g_{B-L}^2}{g_R^2+4g_{B-L}^2} \simeq 0.302 \,.
\eeq
The current Planck bound $T_{D} \gtrsim 600\,\MeV$ implies an upper limit on the $W_R$ and $Z'$ mediated scattering rate, 
which can be translated into a lower bound on $v_R$. 
By numerically solving the equation, 
\beq
\Gamma_{\nu_R}(T) \lesssim H(T) ~~~ \mbox{at $T=600\,\MeV$} \,,
\eeq
we obtain 
\beq
v_R \gtrsim 23\,\TeV\,.
\label{eq:vR_bound}
\eeq

\section{Dark matter physics}
\label{sec:DM}

We impose the global symmetries, $Z^L_4 \times Z^R_4 \times U(1)_L$, in our model in addition to the LR gauge symmetry $G_{LR}$, to control the radiative generation of the small neutrino masses.
The model also has a global $Z_2$ symmetry accidentally, under which $\widetilde{L}_{L,R}^\alpha$ and $n^a_{L,R}$ are odd while all other fields are even. 
The lightest of those particles is thus stable and serves as a good DM candidate if it is electrically neutral. 
As we saw in Sec.~\ref{sec:illustration}, fitting the neutrino oscillation date favors the existence of a light neutral scalar and fermion in the spectrum, and our model naturally accommodates a DM candidate. 
In this section, we study constraints on the DM candidates from the relic abundance and direct detection bounds.

In Sec.~\ref{sec:illustration}, 
we have demonstrated the lepton mass generation in two scenarios. 
Since the mass spectrum is hierarchical in either scenario, 
only the light degrees of freedom are relevant to DM physics. 
In the light $h_R$ scenario where $\wt\nu_R^1$ is light, 
DM is either $n^1$ or $\wt\nu_R^1$. 
In this case, the couplings relevant to DM physics is given by 
\begin{equation}
{\cal L}_{h_R} 
= -\lambda^1_i \ol{e^i_L}\,\wt{e}_L^i n_R^1 
-\left\{ \(\lambda^\nu_L\)_{ij}\,\ol{\nu^i_L}\,\wt{\nu}_L^j n_R^1 
+\(\lambda^\nu_R\)_{i1}\,\ol{\nu^i_R}\,\wt{\nu}_R^1 n_L^1\right\} 
+ \hc\,,
\label{eq:L_hR}
\end{equation}
where $\left(\lambda^\nu_L\right)_{ij}=\lambda^1_i (V_{h_L})_{ij}$ and $\left(\lambda^\nu_R\right)_{ij}=\lambda^1_i (V_{h_R})_{ij}$.
As we will discuss in detail below, DM is thermally produced via its couplings to the right-handed neutrinos, while the DM couplings to the charged leptons and left-handed neutrinos are relevant to direct detection. 
In the light $h_L$ scenario where $h_L^1$ is light, 
DM is either $n^1$ or $h_L^1$, and the couplings relevant to DM physics is given by 
\begin{equation}
{\cal L}_{h_L}
= -\lambda^1_i \ol{e^i_L}\,\wt{e}_L^i n_R^1 
-\frac{1}{\sqrt{2}} \left\{ \(\lambda^\nu_L\)_{ij}\,\ol{\nu^i_L}\,h_L^j n_R^1 
+\(\lambda^\nu_R\)_{i1}\,\ol{\nu^i_R}\,h_R^1 n_L^1\right\} 
+ \hc
\label{eq:L_hL}
\end{equation}
In this case, the DM thermal production is achieved with the couplings to the left-handed neutrinos. 
The couplings to the charged leptons make the main contribution to direct detection.  
Note that since the mass mixings between $\wt\nu_L$ and $\wt\nu_R$ are small, we ignore them in our analysis. 
These interactions, Eqs.(\ref{eq:L_hR}) and (\ref{eq:L_hL}), are quite similar to the ones of the lepton portal DM \cite{Bai:2014osa, Chang:2014tea}. 
The phenomenology and parameter space of the lepton portal DM 
have been comprehensively studied in the literature \cite{Kawamura:2020qxo, Okawa:2020jea, Iguro:2022tmr}, and we can apply their findings to our model.

\subsection{Relic abundance}

To fit the lepton mass matrices, $\lambda^1_i$ cannot be arbitrarily small. 
For the coupling values that can reproduce the realistic lepton mass structures, the DM candidate particles frequently scatter with the SM leptons at high temperatures, thereby bringing the former into thermal equilibrium with the SM plasma.
This suggests that our DM candidates can be thermally produced.
Here, we study the conventional thermal freeze-out production for each DM candidate and clarify viable parameter region.

\subsubsection{Bounds from $\nu_R$ thermalization and decoupling}

In the light $h_R$ scenario, DM is expected to be thermally produced through its annihilation into the right-handed neutrinos $\nu_R$. 
In that case, $\nu_R$ should be in thermal equilibrium with the SM plasma until the DM production is completed.
However, the $\nu_R$ thermalization will increase the radiation density in the universe, as discussed in Sec.~\ref{sec:Neff}. 
Namely, the thermalization rate $\Gamma_{\nu_R}(T)$ has to be large enough for $\nu_R$ to keep in the thermal bath during the DM production, while it has to be small enough for $\nu_R$ to decouple from the thermal bath before the plasma temperature cools down to $T=600$\,MeV.
This requirement translates into a lower limit on DM mass, as well as into an upper limit on $v_R$ for a given DM mass.
We shall derive those bounds below.

First, we obtain a lower limit on DM mass by requiring $\nu_R$ to be decoupled from the SM plasma after the freeze-out: 
\beq
T_{\rm fo} \sim m_\DM/20 > T_D \gtrsim 600\,\MeV \Rightarrow m_\DM \gtrsim 10\,\GeV\,.
\eeq
Here, we required $\nu_R$ to have the same temperature as the plasma temperature during the DM production. 
DM could be produced from $\nu_R$ that has a different temperature, which would enable a lighter DM \cite{Borah:2025fkd}. 
We do not consider such a case here. 

Next, the upper limit on $v_R$ is obtained by requiring $\nu_R$ to keep in the thermal bath at the time of the DM freeze-out. 
Solving the thermalization condition, 
\beq
\Gamma_{\nu_R}(T) \gtrsim H(T) ~~~ \mbox{at $T=T_{\rm fo} \simeq m_\DM/20$} \,,
\eeq
we get the upper limit, 
\beq
v_R \lesssim 125\,\TeV \times \(\frac{m_\DM}{100\,\GeV}\)^{3/4} \,,
\label{eq:vR_thermal}
\eeq
for a given DM mass.
Here we considered that the $W_R$ and $Z'$-mediated processes mainly contribute to the $\nu_R$ thermalization. 
This limit should be satisfied only if DM is thermally produced by its annihilation into $\nu_R$. 
In our benchmark scenarios, we take $v_R=25\,\TeV$, 
which allows DM to be as light as 10\,GeV.

\subsubsection{Thermal production}
In the thermal freeze-out scenario, 
DM abundance is determined by solving the Boltzmann equation,
\begin{align}
\frac{dn_{\rm dm}}{dt}+3Hn_{\rm dm}=-\langle\sigma v\rangle_{\text{eff}}\left[n_{\rm dm}^{2}-\left(n_{\rm dm}^{\rm eq}\right)^{2}\right]\,,
\end{align}
where $n_{\rm dm}^{(\rm eq)}$ is the (equilibrium) number density of DM, and 
$\langle\sigma v\rangle_{\text{eff}}$ is the effective DM annihilation cross section, which includes DM pair annihilation and coannihilation contributions \cite{Edsjo:1997bg}. 
It is known that the observed DM abundance is explained when $\vev{\sigma v}_{\rm eff} \simeq 3\times 10^{-26}\,{\rm cm^3/s}$, almost independently of DM mass and spin \cite{Saikawa:2020swg}. 

In our case, the dominant process is DM pair annihilation into the neutrinos.
In the light $h_R$ scenario, DM is either the Dirac fermion $n^1$ or the complex scalar $\wt\nu_R^1$, depending on the mass spectrum.
In either case, the pair annihilation into the right-handed neutrinos dominates the production. 
The cross section in the non-relativistic regime is approximately given by 
\begin{align}
(\sigma v)_{n^1\ol{n}^1\to \nu_{R}^{i}\ol{\nu}_{R}^{j}}\simeq
\frac{\left|(\lambda_R^\nu)_{i1} (\lambda_R^{\nu*})_{j1}\right|^2}{32\pi M_1^2\(1+m_{\wt\nu R1}^2/M_1^2\)^2}\,,
\end{align}
for the Dirac fermion $n^1$, 
and 
\begin{align}
(\sigma v)_{\wt\nu_R^1\wt\nu_R^{1\dagger}\to\nu_{R}^i\ol{\nu}_{R}^j}\simeq \frac{\left|(\lambda_R^\nu)_{i 1} (\lambda_R^{\nu*})_{j1}\right|^2 v^2}{48\pi m_{\wt\nu R1}^2\(1+M_{1}^2/m_{\wt\nu R1}^2\)^2}\,,
\end{align} 
for the complex scalar $\wt\nu_R^1$. 
Here, $v$ denotes the relative velocity of DM.
The cross sections are velocity suppressed in the scalar DM case.
In the light $h_L$ scenario, DM is either the Dirac fermion $n^1$ or the real scalar $h_L^1$, 
which is produced mostly through its pair annihilation into the left-handed neutrinos. 
The annihilation cross section is approximated to 
\begin{align}
(\sigma v)_{n^1\ol{n}^1\to \nu_{L}^{i}\ol{\nu}_{L}^{j}}\simeq
\frac{\left|(\lambda_L^\nu)_{i1} (\lambda_L^{\nu*})_{j1}\right|^2}{128\pi M_1^2\(1+m_{h_L1}^2/M_1^2\)^2} \,,
\end{align}
for the Dirac fermion $n^1$, and 
\begin{align}
{(\sigma v)_{h_L^1 h_L^1\to\nu_{L}^i\ol{\nu}_{L}^j}}\simeq
\frac{\left|(\lambda_L^\nu)_{i1} (\lambda_L^{\nu*})_{j1}\right|^2 v^4}{60\pi m_{h_L1}^2\(1+M_1^2/m_{h_L1}^2\)^4} \,,
\end{align}
for the real scalar $h_L^1$. 
There is additional contribution from DM coannihilation processes involving several heavier particles, such as Dirac fermions $n^{2,3}$ and a couple of the charged and neutral scalars. 
However, these particles are supposed to be much heavier than $n^1$, $\wt{\nu}_R^1$ and $h_L^1$, and their contribution is negligible unless DM mass is around the EW scale. 
In our analysis, the Boltzmann equation is numerically solved using \texttt{micromegas\_5.2.13} \cite{Belanger:2020gnr}, which automatically includes all relevant annihilation and coannihilation processes.

\begin{figure}[t]
    \centering
    \begin{subfigure}[b]{0.50\textwidth}
        \includegraphics[width=\textwidth]{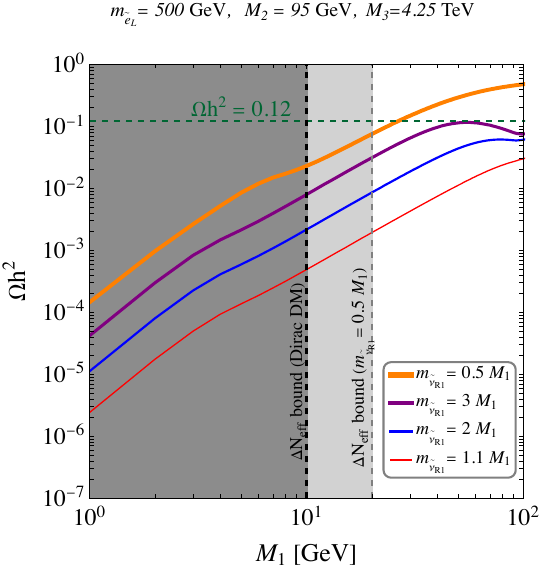}
        \label{fig:model11}
    \end{subfigure}
    \hfill
    \begin{subfigure}[b]{0.49\textwidth}
        \includegraphics[width=\textwidth]{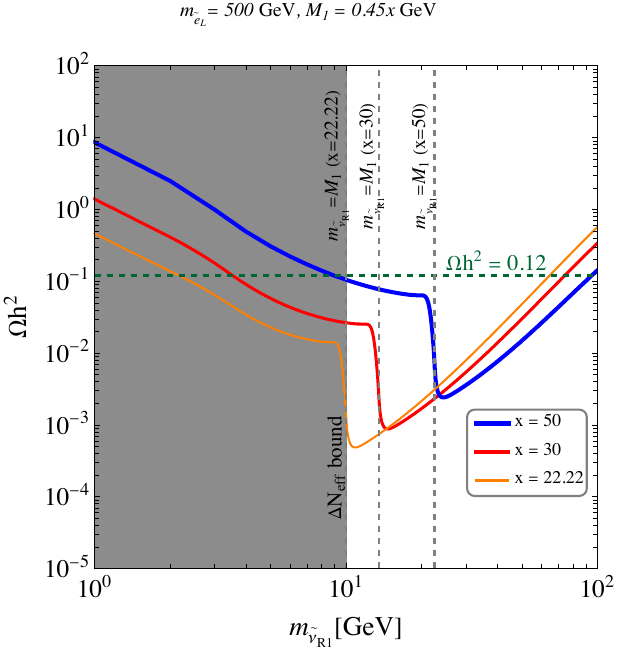}
        \label{fig:model12}
    \end{subfigure}
    \caption{
    The predictions for the DM thermal relic abundance in the light $h_R$ scenario for various values of $m_{\wt\nu R1}^{}/M_1$ (left) and $x=M_1/(0.45\,\GeV)$ (right). 
    The scalar mixing angle is fixed to $\theta=0.5$ and the DM couplings to $\lambda_2^1=\lambda_3^1=\lambda_3^2=1$.
    The value of $\lambda_1^1$ is determined by fitting the electron mass and roughly scales as $\lambda_1^1 \simeq1/\sqrt{x}$.    
    The dark green dashed line represents the observed DM abundance. 
    The $\Delta N_{\text{eff}}$ bound excludes $m_{\rm DM}\lesssim10\,\GeV$, which means that $M_1\lesssim10\,\GeV$ is excluded for Dirac DM (gray) and $M_1\lesssim20\,\GeV$ is excluded for complex scalar DM with $m_{\wt\nu R1}^{}/M_1=0.5$ (light gray).
    }
    \label{fig:relic-model1}
\end{figure}

In Fig.\,\ref{fig:relic-model1}, we show the predictions for the DM thermal relic abundance in the light $h_R$ scenario as a function of $M_1$ (left) and of $m_{\wt\nu_{R1}}$ (right). 
In both plots, we assume the mass spectrum in Eq.~(\ref{eq:inputs_lighthR}), the coupling structure in Eq.~(\ref{eq:Yukawa-lambda}) and the scalar mixing matrices in Eq.~(\ref{eq:mixing_model1}), with $\theta=0.5$ and $\lambda_2^1=\lambda_3^1=\lambda_3^2=1$.
The size of the coupling $\lambda_1^1$ is determined by fitting the electron mass and roughly scales as $\lambda_1^1\simeq1/\sqrt{x}$, where $x:=M_1/(0.45\,\GeV)$.

In the left panel of Fig.\,\ref{fig:relic-model1}, 
DM is the complex scalar $\wt\nu_R^1$ for $m_{\wt\nu R1}^{}/M_1=0.5$ (orange) and the Dirac fermion $n^1$ for $m_{\wt\nu R1}^{}/M_1=3$ (purple), 2 (blue) and 1.1 (red). 
In general, the DM abundance is small in the light $M_1$ region, since a smaller $M_1$ leads to a larger DM coupling $\lambda^1_1$, which in turn increases the annihilation cross section.
Besides, a larger mass splitting leads to a larger DM abundance due to the fact that the annihilation cross section decreases as the mass splitting increases. 
In the scalar DM case, the annihilation cross section receives the velocity suppression at the freeze-out, which leads to the larger DM abundance than in the Dirac case. 
Note that in the Dirac DM case, $n^1$ becomes heavier than $n^2$ for $M_1>95\,\GeV$, so the latter is DM in that mass region. 
As shown in Sec.~\ref{sec:Neff}, the $\Delta N_{\text{eff}}$ bound limits the DM mass to be heavier than about 10\,GeV. 
This bound translates into $M_1\gtrsim10\,\GeV$ for the Dirac DM (gray) and $M_1\gtrsim20\,\GeV$ for complex scalar DM with $m_{\wt\nu R1}^{}/M_1=0.5$ (light gray).
The correct DM abundance is obtained at $M_1\simeq25\,\GeV$ with $m_{\wt\nu R1}^{}=0.5M_1$ and $M_1\simeq50\,\GeV$ with $m_{\wt\nu R1}^{}=3M_1$.

In the right panel of Fig.\,\ref{fig:relic-model1}, 
we show the DM thermal relic abundance as a function of $m_{\wt\nu R1}^{}$, with various values of $x=M_1/(0.45\,\GeV)$.
The vertical dashed lines indicate the boundaries at which the DM candidate changes. 
For each value of $x$, the complex scalar $\wt\nu_R^1$ serves as DM on the left side of the dashed line, while the Dirac fermion $n^1$ becomes DM on the right side. 
When $\wt\nu_R^1$ is DM, the pair annihilation $\wt\nu_R^1 (\wt\nu_R^1)^\dagger \to \nu_R^{} \ol{\nu}_R^{}$ dominates the production basically. 
However, as $m_{\wt\nu R1}$ gets heavy and becomes comparable with $M_1$, the coannihilation process $n^1\ol{n}^1 \to \nu_R^{} \ol{\nu}_R^{}$ becomes efficient and starts to dominate the production, which causes a sharp drop of the DM abundance.
When $n^1$ is DM, the pair annihilation $n^1\ol{n}^1 \to \nu_R^{} \ol{\nu}_R^{}$ is the leading production process. 
Altogether, the DM mass range from 10\,GeV to 100\,GeV is favored in the $h_R$ scenario in both Dirac and scalar DM cases.

\begin{figure}[t]
    \centering
    \begin{subfigure}[b]{0.48\textwidth}
        \includegraphics[width=\textwidth]{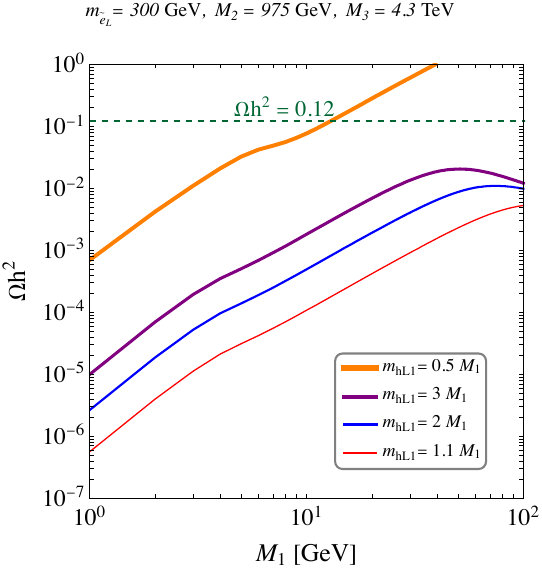}
        \label{fig:model21}
    \end{subfigure}
    \hfill
    \begin{subfigure}[b]{0.48\textwidth}
        \includegraphics[width=\textwidth]{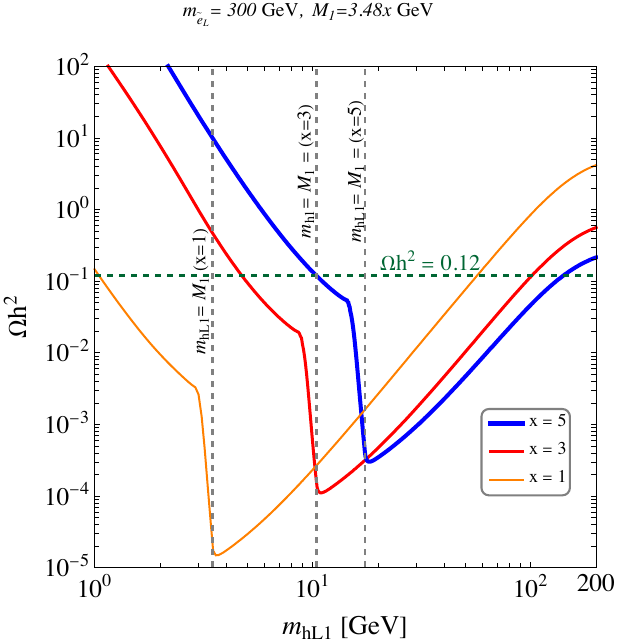}
        \label{fig:model22}
    \end{subfigure}
    \caption{
    The predictions for the DM thermal relic abundance in the light $h_L$ scenario for various values of $m_{h_L1}^{}/M_1$ (left) and $x=M_1/(3.48\,\GeV)$ (right). 
    We consider the small mass splitting $\sqrt{\Delta m_R^2}=300\,\GeV$ between $\wt e_R^1$ and $\wt e_R^{2,3}$, and take $\theta=1$ and $\lambda_2^1=\lambda_3^1=\lambda_3^2=1$.
    The value of $\lambda_1^1$ is determined by fitting the electron mass and roughly scales as $\lambda_1^1 \simeq1/\sqrt{x}$.
    }
    \label{fig:relic-model2}
\end{figure}

Similarly, we show in Fig.\,\ref{fig:relic-model2} the predictions for the the DM abundance in the light $h_L$ scenario as a function of $M_1$ (left) and of $m_{h_L1}$ (right). 
DM is either the Dirac fermion $n^1$ or the real scalar $h_L^1$.
In both left and right plots, we assume the mass spectrum in Eq.~(\ref{eq:inputs_lighthL}), the coupling structure in Eq.~(\ref{eq:Yukawa-lambda}) and the scalar mixing matrices in Eq.~(\ref{eq;model3_mixing}), with $\theta=1$ and $\lambda_2^1=\lambda_3^1=\lambda_3^2=1$.
The value of $\lambda_1^1$ is determined by fitting the electron mass and roughly scales as $\lambda_1^1\simeq1/\sqrt{x}$, where $x:=M_1/(3.48\,\GeV)$. 
It should be noted that we take a small mass splitting $\sqrt{\Delta m_R^2}=300$\,GeV between $\wt e_R^1$ and $\wt e_R^{2,3}$, which is required to fit the neutrino oscillation data. 
This mass splitting is, however, too small to affect the DM physics.

It can be seen in Fig.\,\ref{fig:relic-model2} that the DM thermal relic abundance exhibits qualitatively similar behaviors to the ones in the light $h_R$ scenario. 
Nonetheless, a few differences can be observed. 
First, in the real scalar DM case (with $m_{h_L1}=0.5M_1$), the pair annihilation cross section receives the $v^4$ velocity suppression, in contrast to $v^2$ in the complex scalar DM case. 
This stronger suppression results in a larger DM abundance for a fixed DM mass. 
Second, for a given $M_1$, the value of $\lambda_1^1$ is larger than the one in the light $h_R$ scenario, because $\wt e_R$ is heavier. 
The large coupling therefore leads to the suppressed DM abundance, compared with the light $h_R$ scenario. 
Lastly, since DM is produced mainly from its pair annihilation into $\nu_{L}^i\ol{\nu}_L^j$, the $\Delta N_{\text{eff}}$ bound discussed in Sec.~\ref{sec:Neff} does not apply for this scenario. 
As a result, the DM mass range below 10\,GeV is viable in the light $h_L$ scenario.

\subsection{Direct detection}

DM direct detection using nuclear recoils is a powerful experimental method to search for the weak-scale particle DM.
The current direct detection bounds are so strong that DM coupled predominantly to the SM leptons can be excluded. 
In this section, we evaluate the spin-independent (SI) DM-nucleon scattering cross section, which is most stringently constrained by the experiments, and check if our model is consistent with the current experimental bounds. 

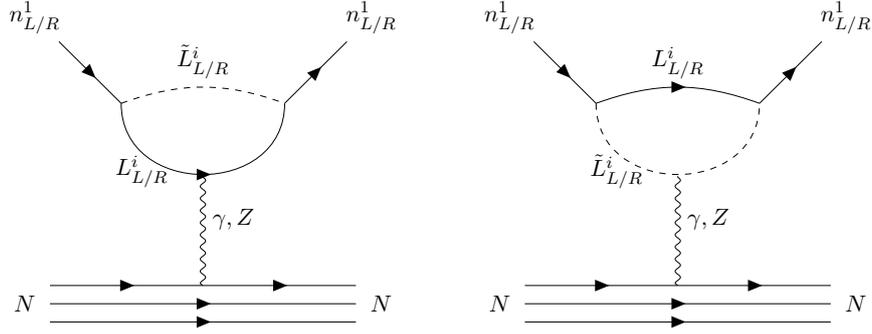
\begin{figure}[t]
\centering
\begin{minipage}{0.4\textwidth}
\centering
\resizebox{0.9\textwidth}{!}{
\begin{tikzpicture}
  \begin{feynman}
    \vertex (a) {$n_{L/R}^1$};
    \vertex [right=5.5cm of a] (c) {$n_{L/R}^1$};

    \vertex [below right=2cm of a] (v1);
    \vertex [below left=2cm of c] (v2);
    \vertex at ($(v1)!0.5!(v2)$) (vtop);

    \vertex [below=1.2cm of vtop] (vmid);
    \vertex [below=1.8cm of vmid] (vbot);

    \vertex [left=2.5cm of vbot] (n1);
    \vertex [right=2.5cm of vbot] (n2);
    \vertex [below=0.3cm of n1] (n3);
    \vertex [below=0.3cm of n2] (n4);
    \vertex [below=0.3cm of n3] (n5);
    \vertex [below=0.3cm of n4] (n6);

    \diagram* {
      (a) -- [fermion] (v1),
      (v2) -- [fermion] (c),

      (v1) -- [scalar, bend left=20, edge label={$\tilde{L}^i_{L/R}$}] (v2),
      (v1) -- [fermion, half right] (v2),

      (vmid) -- [photon] (vbot),

      (n1) -- [fermion] (vbot) -- [fermion] (n2),
      (n3) -- [fermion] (n4),
      (n5) -- [fermion] (n6),
    };

    \node at (n3) [left=0.1cm] {$N$};
    \node at (n4) [right=0.1cm] {$N$};
    \node at ($(vmid)+(0.5cm,-0.7cm)$) {$\gamma, Z$};
    \node at ($(vmid)+(-1cm,+0.08cm)$) {$L_{L/R}^i$};
  \end{feynman}
\end{tikzpicture}
}
\end{minipage}
\begin{minipage}{0.4\textwidth}
\centering
\resizebox{0.9\textwidth}{!}{
\begin{tikzpicture}
  \begin{feynman}
    \vertex (a) {$n_{L/R}^1$};
    \vertex [right=5.5cm of a] (c) {$n_{L/R}^1$};

    \vertex [below right=2cm of a] (v1);
    \vertex [below left=2cm of c] (v2);
    \vertex at ($(v1)!0.5!(v2)$) (vtop);

    \vertex [below=1.2cm of vtop] (vmid);
    \vertex [below=1.8cm of vmid] (vbot);

    \vertex [left=2.5cm of vbot] (n1);
    \vertex [right=2.5cm of vbot] (n2);
    \vertex [below=0.3cm of n1] (n3);
    \vertex [below=0.3cm of n2] (n4);
    \vertex [below=0.3cm of n3] (n5);
    \vertex [below=0.3cm of n4] (n6);

    \diagram* {
      (a) -- [fermion] (v1),
      (v2) -- [fermion] (c),

      (v1) -- [fermion, bend left=20, edge label={$L^i_{L/R}$}] (v2),
      (v1) -- [scalar, half right] (v2),

      (vmid) -- [photon] (vbot),

      (n1) -- [fermion] (vbot) -- [fermion] (n2),
      (n3) -- [fermion] (n4),
      (n5) -- [fermion] (n6),
    };

    \node at (n3) [left=0.1cm] {$N$};
    \node at (n4) [right=0.1cm] {$N$};
    \node at ($(vmid)+(0.5cm,-0.7cm)$) {$\gamma, Z$};
    \node at ($(vmid)+(-1cm,+0.08cm)$) {$\tilde{L}_{L/R}^i$};
  \end{feynman}
\end{tikzpicture}
}
\end{minipage}
\vspace{0.3cm}
\caption{The leading one-loop diagrams for elastic scattering of the Dirac DM with a nucleon ($N$).}
\label{fig:feynman_loop}
\end{figure}

If DM is the Dirac fermion $n^1$, the DM-nucleon scattering only arises at the one-loop level through its couplings to the SM leptons. 
This case is well studied in the context of the lepton portal DM models, and we can directly use the results in \cite{Kawamura:2020qxo, Iguro:2022tmr}. 
The leading contribution is from one-loop photon and $Z$-mediated processes shown in Fig.\,\ref{fig:feynman_loop}. 
The SI DM-nucleon scattering cross section is then given by 
\begin{align}
\sigma_{\rm SI}^{\rm Dirac} = \frac{m_n^2M_1^2}{\pi(m_n+M_1)^2}\left(\frac{ZC_{V,p}+(A-Z)C_{V,n}}{A}\right)^2
\end{align}
where $m_n$ is the nucleon mass, $Z$ and $A$ are the atomic number and mass of a target nucleus, and 
\beq
C_{V,N}=C_{V,N}^{\gamma}+C_{V,N}^{Z} \quad\mbox{with $N=p,n$}\,.
\eeq
Here, $C_{V,N}^{\gamma}$ and $C_{V,N}^{Z}$ denote the photon and $Z$-penguin contribution, given by 
\beq
C_{V,N}^{\gamma} & \simeq \frac{e^2 Q_N}{96\pi^2}
\sum_{i} \frac{\left|\lambda_{i}^{1}\right|^2}{m_{\wt eL}^2}
\left(\frac{3}{2}+\log\frac{m_i^2}{m_{\wt eL}^2}\right) \,, \\
C_{V,N}^Z & \simeq \frac{g^2g_{V,N}^{}}{128\pi^2m_Z^2c_W^2} \sum_i|(\lambda_L^\nu)_{i1}|^2
\left(1-\frac{1}{2}\frac{m_{A_Li}^2+m_{h_L1}^2}{m_{A_Li}^2-m_{h_L1}^2} \log\frac{m_{A_Li}^2}{m_{h_L1}^2} \right) \,,\label{eq:Zpen2}
\eeq
where $Q_N$ is the electric charge of the nucleon, $g_{V,p}^{}=1/2-2s_W^2$ and $g_{V,n}^{}=-1/2$. 
To obtain the above equations, the $m_\chi \to 0$ limit is taken. 
We note that all $A_L^i$ have a common mass $m_{ALi}^{}=m_{\wt\nu L23}^{}$ in our benchmark study.
There is also the $\wt e_R^i$ contribution to $C_{V,N}^\gamma$, which is obtained simply by $m_{\wt eR} \to m_{\wt eL}$. 
Given $m_{\wt eL} \ll m_{\wt eR}$, however, this contribution is negligible in our model.
The $Z$-penguin contribution $C_{V,N}^Z$ is important only if there is a large mass splitting between $h_L^1$ and $A_L^i$, namely in the light $h_L^1$ scenario \cite{Iguro:2022tmr}. 

If DM is the complex scalar $\wt\nu_R^1$, 
both photon and $Z$-penguin processes are vanishing, since $\wt\nu_R^1$ does not directly couple to the charged leptons nor the left-handed neutrinos.
The leading contribution arises from the tree-level $Z'$-mediated process. 
Then the SI cross section is given by 
\begin{align}
\sigma_{\rm SI}^{\rm complex} = \frac{m_n^2 m_{\wt\nu R1}^2}{\pi(m_n+m_{\wt\nu R1})^2}\left(\frac{ZC_{V,p}^{Z'}+(A-Z)C_{V,n}^{Z'}}{A}\right)^2 \,,
\end{align}
where 
\beq
C_{V,N}^{Z'}=-\frac{8 G_F}{\sqrt{2}}\frac{v_L^2}{v_R^2}\,T_{R,\tilde{\nu}_R}^3 g_{V,N}^{Z'}\,,
\eeq
with $T_{R,\tilde{\nu}_R}^3=1/2$. 
The $Z'$-nucleon couplings are given by $g_{V,p}^{Z'}=-3s_R^2/4+1/4$ and $g_{V,n}^{Z'}=-s_R^2/4-1/4$ with $s_R^2\simeq 0.302$.

If DM is the real scalar $h_L^1$, 
the DM-nucleon scattering induced from the DM-lepton couplings are very suppressed and negligible \cite{Kawamura:2020qxo}. 
The leading process happens in a similar way to the inert doublet DM, stemming mainly from the Higgs portal interaction and the loop corrections with the weak interactions \cite{Abe:2015rja, Higuchi:2023kbt}.
The SI DM-neutron scattering cross section is given by \cite{Abe:2015rja} 
\begin{equation}
\sigma_{\rm SI}^{\rm real} = \frac{|\lam_{345}+\delta\lam_n|^2 m_n^4 f_n^2}{4\pi (m_{h_L1}+m_n)^2 m_h^4} \,,
\label{eq:sigmaSI}
\end{equation}
where $m_h$ is the Higgs boson mass, 
$f_n\simeq0.287$, and 
$\lam_{345}\simeq\lambda_{11}^e+\lambda^\nu_{11}+\lambda^s_{11}$ represents the coupling between $h_L^1$ and the Higgs boson, which is a free parameter in our analysis. 
The loop corrections are rendered in $\delta\lambda_n$, which is approximately given in the light $h_L^1$ regime by \cite{Higuchi:2023kbt}
\beq
\delta \lambda_n & \simeq 
    -0.00199 
    +(1.18\,m_{h_L1} -6.30\,m_{\wt eL}
    -4.46\times10^{-3}\,m_{\wt eL}^2)\times10^{-6}  
      \nonumber\\
    & \quad 
    + \lambda_{\wt L_L} \left( 
    0.00164
    + 2.57/m_{\wt eL}^2 
    + 5.76\times10^3/m_{\wt eL}^4 
    \right) \,,
    \label{eq:dellam_approx}
\eeq
where all masses are in GeV and $\lambda_{\wt L_L}$ is the self-coupling of $\wt{L}_L$.
This cross section is independent of $\lambda^a_i$, and can be arbitrarily suppressed by appropriately choosing the value of $\lambda_{\wt L_L}$. 
Thus, we do not consider this case further.

\begin{figure}[t]
    \centering
    \includegraphics[width=0.55\textwidth]{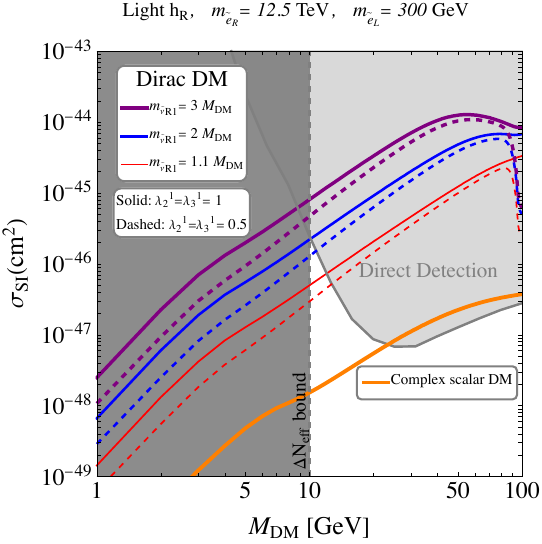}
    \caption{
    The SI DM-nucleon scattering cross section $\sigma_{\rm SI}$ in the light $h_R$ scenario. 
    The SI cross section, multiplied by a factor of $(\Omega h^2)_{\rm thermal}/0.12$, is shown, to reflect the fact that the produced DM abundance constitutes only a fraction of the total DM abundance in our model.
    The purple, blue and red lines correspond to the Dirac DM with $\lambda_2^1=\lambda_3^1=1$ (solid) and $\lambda_2^1=\lambda_3^1=0.5$ (dashed), while the orange line corresponds to the complex scalar DM. 
    The light gray region is excluded by the current direct detection bounds. 
    The gray region is not consistent with the $\Delta N_\eff$ bound.
    }
    \label{fig:direct}
\end{figure}

\begin{figure}[t]
    \centering
    \includegraphics[width=0.55\textwidth]{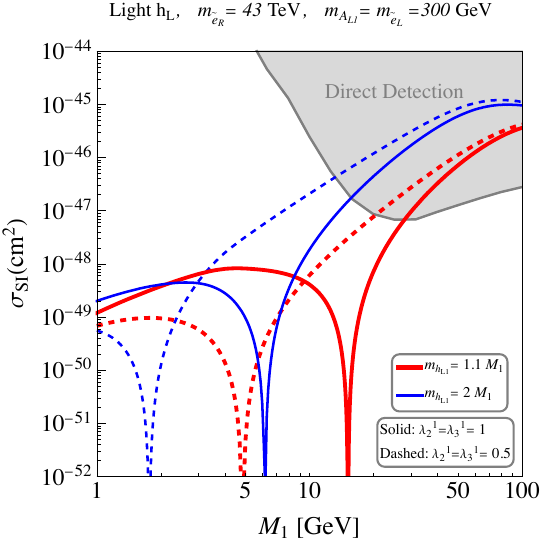}
    \caption{
    The SI DM-nucleon scattering cross section in the light $h_L$ scenario for Dirac DM. 
    We take $m_{h_L1}=1.1M_1$ (red) and $2M_1$ (blue) with $\lambda^1_2=\lambda^1_3=1$ (solid) and $\lambda^1_2=\lambda^1_3=0.5$ (dashed).
    }
    \label{fig:DD2}
\end{figure}

Let us now look at the direct detection bounds in our model.
To compare with the experimental bounds, we rescale the SI cross section with $(\Omega h^2)_{\rm thermal}/(\Omega h^2)_{\rm obs}$, since in a large part of the parameter space, the produced DM abundance is much below the observed value $(\Omega h^2)_{\rm obs}\simeq0.12$. 

In Fig.~\ref{fig:direct} we show the SI cross section in the light $h_R$ scenario with $\lambda_2^1=\lambda_3^1=1$ (solid) and $\lambda_2^1=\lambda_3^1=0.5$ (dashed). 
For consistency with the neutrino mass generation, we take $m_{\wt eR}=12.5\,\TeV$, $m_{\wt eL}=300\,\GeV$ and $\theta=0.5$.
The results for Dirac DM are shown with $m_{\wt\nu_R1}=3M_{\DM}$ (purple), $2M_{\DM}$ (blue) and $1.1M_{\DM}$ (red), where $M_\DM=M_1$ denotes the mass of DM. 
The orange line shows the result for the complex scalar DM. 
Combined the current direct detection limits with the $N_{\text{eff}}$ bound, the allowed DM mass range is restricted to a narrow window $10\,\GeV \lesssim M_\DM \lesssim 15\,\GeV$ for Dirac DM, whereas $10\,\GeV \lesssim M_\DM \lesssim 20\,\GeV$ for complex scalar DM.

Next we investigate the direct detection bounds in the light $h_L$ scenario. 
We only consider the Dirac DM case as mentioned above. 
The leading contribution to the SI cross section arises from either photon or $Z$-penguin diagrams, depending on the choice of the model parameters. 
For $m_i\ll m_{\wt eL}$ and $m_{h_L1}\ll m_{A_Li}$, which we are interested in, the penguin contributions are approximately given by
\beq
C_{V,p} &\simeq C_{V,p}^\gamma \simeq \frac{e^2}{96\pi^2} \sum_{i} \frac{\left|\lambda_{i}^{1}\right|^2}{m_{\wt eL}^2} \log\frac{m_i^2}{m_{\wt eL}^2} \,,\\
C_{V,n} &= C_{V,n}^Z \simeq \frac{e^2}{512\pi^2m_Z^2c_W^2s_W^2} \sum_i|(\lambda_L^\nu)_{i1}|^2 \log\frac{m_{A_Li}^2}{m_{h_L1}^2} \,.
\eeq
Namely, $C_{V,p}$ is negative while $C_{V,n}$ is positive. 
This leads to a potential cancellation in the SI cross section, which is clearly observed in Fig.~\ref{fig:DD2}. 
The cancellation occurs accidentally since the parameter dependence of $C_{V,p}$ and $C_{V,n}$ is completely different. 
We note that the cancellation point depends not only on the model parameters but also on the target nucleus used in the direct detection experiments. 
In Fig.~\ref{fig:DD2}, we combine the bounds from the LZ \cite{LZ:2022lsv} and PandaX-4T \cite{PandaX-4T:2021bab} experiments, which use xenon as a target. 
Therefore, we assume $Z=54$ and $A=131$ in the analysis.
In contrast, DEAP-3600 \cite{DEAP-3600:2017ker, DEAP-3600:2017uua} and DarkSide \cite{DarkSide-50:2022qzh, DarkSide-20k:2017zyg} experiments use argon as a target nucleus, for which we should instead take $Z=18$ and $A=40$. 
Combining different experiments using different target materials would be important to fully explore our model parameter space, including the blind spot caused by the photon and $Z$-penguin cancellation.

\section{Summary and discussion}
\label{sec:Summary}

The LR symmetric extensions are an attractive candidate for physics beyond the SM.
In this paper, we have studied the neutrino mass generation and DM physics in the LR symmetric model. 
The model builds on the $G_{LR} = SU(3)_c \times SU(2)_L \times SU(2)_R \times U(1)_{B-L}$ gauge symmetry. 
The quark sector of our model is similar to the one presented in Refs. \cite{Bolton:2019bou,Babu:2022ikf}, while the lepton sector is modified.
Instead of the vectorlike lepton fields, we newly introduce a couple of the gauge singlet fermions $n^a_{L,R}$, $SU(2)_L$ doublet scalars $\wt{L}_L^\alpha$ and $SU(2)_R$ doublet scalars $\wt{L}_R^\alpha$. 
These are coupled to the lepton doublets $L_{L,R}^i$, which contains the SM charged leptons and left-handed and right-handed neutrinos. 
The tiny neutrino masses can be generated from the three-loop diagrams via those couplings.
We have identified two specific scenarios that can successfully generate the realistic structures of the charged lepton and neutrino mass matrices. 
In one case, the complex scalar $\wt\nu_R^1$, which is the lightest neutral component of $\wt L_R^\alpha$, has a mass lighter than the EW scale. 
In the other case, the real scalar $h_L^1$, that is the lightest neutral component of $\wt L_L^\alpha$, has to reside below the EW scale.
In either scenario, the remaining scalars should be much heavier than the lightest one to reproduce the neutrino oscillation parameters.
See Sec.~\ref{sec:illustration} for further details. 

Our model possesses an accidental global $Z_2$ symmetry, 
under which only $n^a_{L,R}$ and $\wt L_{L,R}^\alpha$ are odd 
while all other fields are even. 
As a result, the lightest neutral particle among them is a good DM candidate. 
There appear three DM candidates depending on the mass spectrum of $n^a_{L,R}$ and $\wt L_{L,R}^\alpha$: 
(i) the Dirac fermion $n^1$, 
(ii) the complex scalar $\wt\nu_R^1$, 
and (iii) the real scalar $h_L^1$.
We have studied the phenomenology of those DM candidates with several benchmark parameter sets that can reproduce the charged lepton and neutrino mass matrix structures. 
Then, we have found that all DM candidates can be thermally produced with the correct relic abundance. 
In the light $h_R$ scenario, DM annihilation into right-handed neutrinos dominates the production. 
In this case, the $\Delta N_\eff$ bound from the thermalization of right-handed neutrinos limits the allowed DM mass range to be $M_\DM \gtrsim 10\,\GeV$, in which region the correct DM abundance is achieved. 
In the light $h_L$ scenario, the $\Delta N_\eff$ bound does not apply, 
thereby extending the allowed DM mass to the sub-GeV region. 
Regarding the DM direct detection, 
the elastic DM-nucleon scattering is induced mainly from one-loop photon and $Z$-penguin diagrams or tree-level $Z'$ exchange. 
The current direct detection experiments put very strong bounds, especially on the high mass region.
We have found that only a limit window of $M_\DM \lesssim 15\,\GeV$ is consistent with the experimental results, 
unless there is a significant cancellation between the photon and $Z$-penguin contributions. 
See Figs.~\ref{fig:relic-model1}, \ref{fig:relic-model2}, \ref{fig:direct} and \ref{fig:DD2} for the relic abundance and direct detection bounds in the benchmark scenarios.

In this paper, we have only considered two specific scenarios for the extra particle spectrum to simplify the situation and to ease the analysis. 
An unexplored but viable parameter space might exist. 
The entire survey of the model parameter space is still in progress.

\section*{Acknowledgements}
We thank Syuhei Iguro for the collaboration at the early stage of this work and giving useful comments on the manuscript.
This work is supported in part by the Grant-in-Aid for Scientific Research from the MEXT, Japan, No.\,24K07031 (Y.\,O.) and by the JSPS KAKENHI Grant Numbers 22K21350 and 25K17401 (S.O.).
The work of S.O. is also supported by an appointment to the JRG Program at the APCTP through the Science and Technology Promotion Fund and Lottery Fund of the Korean Government and by the Korean Local Governments -- Gyeongsangbuk-do Province and Pohang City.
The work of K.W. is supported by the China Scholarship Council (CSC), the grants PID2022-126224NB-C21 and 2021-SGR-249 (Generalitat de Catalunya).

\appendix

\section{The evaluation of $\kappa_\nu^{\rm equiv}$}
\label{appendix}

\subsection{The contribution of diagram (A)}
We adopt the Feynman-`t Hooft gauge,\footnote{In our model, the Goldstone bosons that come from $H_{L/R}$ do not appear in the diagram (C). }
and derive the following description:
\beq
\label{kappanuA}
i\left (\kappa_\nu^{\rm equiv} \right )_{\alpha \beta} =& \frac{ g^2 }{ 2} \left (V_{\widetilde e L} \right )_{\alpha \alpha^\prime} \widetilde \kappa_{\alpha^\prime \beta^\prime} \left (V^\dagger_{\widetilde e R} \right )_{\beta^\prime \beta}  \nonumber \\
&\times \int \frac{ d^n k}{(2 \pi)^n} \frac{ k^2\Delta M^2_{WW_R} (k^2)}{(k^2 -M^2_{W_R}) (k^2 -M^2_{W}) (k^2 -m^2_{\widetilde e_R \beta^\prime}) (k^2 -m^2_{\widetilde e_L \alpha^\prime})}
\eeq
where $g$ denotes the gauge couplings of $SU(2)_L$ and $SU(2)_R$.
$ M^2_{W W_R}$ corresponds to the mass mixing between $W$ and $W_R$ generated by the colored fermion loop. It is given by
\beq
M^2_{W W_R}(k^2)=&\frac{ 3 g^2 }{ (4 \pi)^2} \left \{ \left ( U_{tR} \right )_{21}  m_t \left ( U^\dagger_{tL} \right )_{11} \right \} \left \{ \left ( U_{bL} \right )_{11} m_b \left ( U^\dagger_{bR} \right )_{12} \right \}  \nonumber \\
& \times\int^1_0 dx \, \log \left \{ \frac{(m^2_t x +m^2_b (1-x)-k^2 x(1-x))(m^2_T x +m^2_B (1-x)-k^2 x(1-x))}{(m^2_t x +m^2_B (1-x)-k^2 x(1-x)) (m^2_T x +m^2_b (1-x)-k^2 x(1-x))} \right \},
\eeq
where the following relations are used:
\beq
\left ( U_{tL} \right )_{11} m_t \left ( U^\dagger_{tR} \right )_{12}+\left ( U_{tL} \right )_{12} m_T \left ( U^\dagger_{tR} \right )_{22}&=0, \\
\left ( U_{bL} \right )_{11} m_b \left ( U^\dagger_{bR} \right )_{12}+\left ( U_{bL} \right )_{12} m_B \left ( U^\dagger_{bR} \right )_{22}&=0.
\eeq
Based on the discussion in Sec. \ref{sec:quark}, $\kappa_\nu$ is estimated as
\beq
\left (\kappa_\nu^{\rm equiv} \right )_{\alpha \beta} \approx  \kappa_{\alpha \beta} \frac{g^4}{(4 \pi)^4}  \left ( U_{tR} \right )_{21}   \times {\cal O} \left (10^{-6} \right ),
\eeq
when $m_{\widetilde e_R \alpha}$, $m_T$ and $m_B$ are ${\cal O} (v_R)$ and $m_{\widetilde e_L \alpha}$ is ${\cal O} (v_L)$.

We note that there are diagrams where $W_L$ and/or $W_R$ are replaced by $G^\pm_L$ and/or $G^\pm_R$ in the Diagram (A). 
Those contributions to $\kappa_\nu^{\rm equiv}$ are linear to $\kappa v_L$ or $(\kappa v_L)^2$, so that 
they are expected to be suppressed by $v_L/v_R$, when $m_{\widetilde e_R \alpha}$, $m_T$, $m_B={\cal O}(v_R)$.
In addition, they are also suppressed by $ \left ( U_{tR} \right )_{21} $ originated from $m^U_{33}$.
Eventually, we conclude that the contribution of the diagram (A) is small.

\subsection{The contribution of diagram (B)}

$\kappa_\nu^{\rm equiv}$ is effectively generated via the Goldstone mixing. 
For simplicity, we take the limit that $U_{t_L}$ and $U_{b_L}$ are diagonal and light quark masses, $m_t$ and $m_b$, in the loops are vanishing, in our analysis. Note that we remain only the mass mixing between $t_L$ and $T_R$ ($b_L$ and $B_R$), that is linear to $v_L$.
The $\kappa_\nu^{\rm equiv}$ generated by the diagram (B) is evaluated as
\beq
\label{kappanuC}
iv_L v_R\left (\kappa_\nu \right )_{\alpha \beta} =& \, \kappa_{\alpha \beta}  \int \frac{ d^n k}{(2 \pi)^n} \frac{ \Delta Z_{G_L G_R} (k^2)}{(k^2 -M^2_{W_R}) (k^2 -M^2_{W}) }
\eeq
where
\beq
\Delta Z_{G_L G_R} (k^2)=& \sum^2_{i=1} \frac{3v_L}{8 \pi^2} |y_D|^2 y_U \,  m_B \left ( U^\dagger_{bR} \right )_{21}   \left ( U_{tR} \right )_{1i} \left ( U^\dagger_{tR} \right )_{i2}   \nonumber \\
& \times \left \{   \int^1_0 dx \int^1_0 dy \left ( \frac{k^2 x y^2}{ m^2_{t_i} xy +m^2_B (1-y) - (1-xy)xyk^2} \right ) \right \} \nonumber \\
&+ \sum^2_{i=1}\frac{3v_L}{8 \pi^2} |y_U|^2 y_D m_T  \left ( U_{tR} \right )_{12}  \left ( U_{bR} \right )_{2i} \left ( U^\dagger_{bR} \right )_{i1}    \nonumber \\
& \times \left \{  \int^1_0 dx \int^1_0 dy \left ( \frac{k^2 y (1-y)}{ m^2_{T} xy +m^2_{b_i} (1-y) - y(1-y)k^2} \right )  \right \},
\eeq
where $(m_{t_1},\, m_{t_2})=(m_t, \, m_T)$ and $(m_{b_1},\, m_{b_2})=(m_b, \, m_B)$. In this description, 
the terms linear to $m_b$ and $m_t$ are ignored.
This contribution is estimated as
\beq
\left (\kappa_\nu^{\rm equiv} \right )_{\alpha \beta} \approx  \kappa_{\alpha \beta} \left \{ |y^D_{33}|^2 y^U_{33} \left ( U^\dagger_{tR} \right )_{12} \times {\cal O} \left (10^{-4} \right )   + |y^U_{33}|^2 y^D_{33} \left ( U_{tR} \right )_{12} \left ( U^\dagger_{bR} \right )_{11}   \times {\cal O} \left (10^{-4} \right ) \right \},
\eeq
when $m_T$ and $m_B$ are ${\cal O} (v_R)$.
We conclude that the diagram (B) contribution is dominant and small off-diaonal elements of $U_{tR} $ are required to realize the neutrino mass matrix.

\subsection{The contribution of diagram (C)}
We estimate the size of the diagram (C) contribution. $\left (\Hat M^{(C)}_\nu \right )_{ij} $ is evaluated as 
\beq
\label{kappanuB}
i\left (\Hat M^{(C)}_\nu \right )_{ij} =&  2 g^2 m^i_{\ell}  \,\delta_{ij} 
\int \frac{ d^n k}{(2 \pi)^n} \frac{ \Delta M^2_{WW_R} (k^2)}{(k^2 -M^2_{W_R}) (k^2 -M^2_{W}) (k^2 - m^i_{\ell}) }
\eeq
where $(m^1_{\ell},\, m^2_{\ell}, \, m^3_{\ell})=(m_e, \, m_\mu, \, m_\tau)$. 
Then, the size is estimated as
\beq
\left (\Hat M^{(C)}_\nu \right )_{ij} \approx \frac{g^4}{(4 \pi)^4}  \left ( U_{tR} \right )_{21}  \,  m^i_{\ell}  \,\delta_{ij} \times {\cal O} \left (10^{-4} \right ) \, \[{\rm GeV} \],
\eeq
when $m_T$ and $m_B$ are ${\cal O} (v_R)$.
This contribution appears in the diagonal elements, as long as the approximation is approved. 
${\Hat M}^{(C)}_\nu $ can be also suppressed by $\left ( U_{tR} \right )_{21}$.

\bibliographystyle{utphys28mod}
\bibliography{refLR}

\providecommand{\href}[2]{#2}\begingroup\raggedright\begin{thebibliography}{10}

\bibitem{Beg:1978mt}
M.~A.~B.~Beg and H.~S.~Tsao, ``{Strong P, T Noninvariances in a Superweak
  Theory},'' \href{https://dx.doi.org/10.1103/PhysRevLett.41.278}{Phys.\  Rev.\
   Lett.\  {\bfseries 41} (1978) 278}.

\bibitem{Mohapatra:1978fy}
R.~N.~Mohapatra and G.~Senjanovic, ``{Natural Suppression of Strong p and t
  Noninvariance},''
  \href{https://dx.doi.org/10.1016/0370-2693(78)90243-5}{Phys.\  Lett.\  B
  {\bfseries 79} (1978) 283--286}.

\bibitem{Babu:1989rb}
K.~S.~Babu and R.~N.~Mohapatra, ``{A Solution to the Strong {CP} Problem
  Without an Axion},''
  \href{https://dx.doi.org/10.1103/PhysRevD.41.1286}{Phys.\  Rev.\  D
  {\bfseries 41} (1990) 1286}.

\bibitem{Barr:1991qx}
S.~M.~Barr, D.~Chang, and G.~Senjanovic, ``{Strong CP problem and parity},''
  \href{https://dx.doi.org/10.1103/PhysRevLett.67.2765}{Phys.\  Rev.\  Lett.\
  {\bfseries 67} (1991) 2765--2768}.

\bibitem{Chakdar:2013tca}
S.~Chakdar, K.~Ghosh, S.~Nandi, and S.~K.~Rai, ``{Collider signatures of mirror
  fermions in the framework of a left-right mirror model},''
  \href{https://dx.doi.org/10.1103/PhysRevD.88.095005}{Phys.\  Rev.\  D
  {\bfseries 88} (2013) 095005} {\ttfamily
  [\href{https://arxiv.org/abs/1305.2641}{arXiv:1305.2641}]}.

\bibitem{DAgnolo:2015uqq}
R.~T.~D'Agnolo and A.~Hook, ``{Finding the Strong CP problem at the LHC},''
  \href{https://dx.doi.org/10.1016/j.physletb.2016.09.061}{Phys.\  Lett.\  B
  {\bfseries 762} (2016) 421--425} {\ttfamily
  [\href{https://arxiv.org/abs/1507.00336}{arXiv:1507.00336}]}.

\bibitem{Hall:2018let}
L.~J.~Hall and K.~Harigaya, ``{Implications of Higgs Discovery for the Strong
  CP Problem and Unification},''
  \href{https://dx.doi.org/10.1007/JHEP10(2018)130}{JHEP {\bfseries 10} (2018)
  130} {\ttfamily [\href{https://arxiv.org/abs/1803.08119}{arXiv:1803.08119}]}.

\bibitem{Craig:2020bnv}
N.~Craig, I.~Garcia~Garcia, G.~Koszegi, and A.~McCune, ``{P not PQ},''
  \href{https://dx.doi.org/10.1007/JHEP09(2021)130}{JHEP {\bfseries 09} (2021)
  130} {\ttfamily [\href{https://arxiv.org/abs/2012.13416}{arXiv:2012.13416}]}.

\bibitem{Pati:1974yy}
J.~C.~Pati and A.~Salam, ``{Lepton Number as the Fourth Color},''
  \href{https://dx.doi.org/10.1103/PhysRevD.10.275}{Phys.\  Rev.\  D {\bfseries
  10} (1974) 275--289}. [Erratum: Phys.Rev.D 11, 703--703 (1975)].

\bibitem{Mohapatra:1974hk}
R.~N.~Mohapatra and J.~C.~Pati, ``{Left-Right Gauge Symmetry and an
  Isoconjugate Model of CP Violation},''
  \href{https://dx.doi.org/10.1103/PhysRevD.11.566}{Phys.\  Rev.\  D {\bfseries
  11} (1975) 566--571}.

\bibitem{Mohapatra:1974gc}
R.~N.~Mohapatra and J.~C.~Pati, ``{A Natural Left-Right Symmetry},''
  \href{https://dx.doi.org/10.1103/PhysRevD.11.2558}{Phys.\  Rev.\  D
  {\bfseries 11} (1975) 2558}.

\bibitem{Senjanovic:1975rk}
G.~Senjanovic and R.~N.~Mohapatra, ``{Exact Left-Right Symmetry and Spontaneous
  Violation of Parity},''
  \href{https://dx.doi.org/10.1103/PhysRevD.12.1502}{Phys.\  Rev.\  D
  {\bfseries 12} (1975) 1502}.

\bibitem{Mohapatra:1980yp}
R.~N.~Mohapatra and G.~Senjanovic, ``{Neutrino Masses and Mixings in Gauge
  Models with Spontaneous Parity Violation},''
  \href{https://dx.doi.org/10.1103/PhysRevD.23.165}{Phys.\  Rev.\  D {\bfseries
  23} (1981) 165}.

\bibitem{Iguro:2021nhf}
S.~Iguro, J.~Kawamura, Y.~Omura, and Y.~Shigekami, ``{Higgs flavor
  phenomenology in a supersymmetric left-right model with parity},''
  \href{https://dx.doi.org/10.1007/JHEP06(2021)125}{JHEP {\bfseries 06} (2021)
  125} {\ttfamily [\href{https://arxiv.org/abs/2103.12712}{arXiv:2103.12712}]}.

\bibitem{Iguro:2018oou}
S.~Iguro, Y.~Muramatsu, Y.~Omura, and Y.~Shigekami, ``{Flavor physics in the
  multi-Higgs doublet models induced by the left-right symmetry},''
  \href{https://dx.doi.org/10.1007/JHEP11(2018)046}{JHEP {\bfseries 11} (2018)
  046} {\ttfamily [\href{https://arxiv.org/abs/1804.07478}{arXiv:1804.07478}]}.

\bibitem{Kuchimanchi:1995rp}
R.~Kuchimanchi, ``{Solution to the strong CP problem: Supersymmetry with
  parity},'' \href{https://dx.doi.org/10.1103/PhysRevLett.76.3486}{Phys.\
  Rev.\  Lett.\  {\bfseries 76} (1996) 3486--3489} {\ttfamily
  [\href{https://arxiv.org/abs/hep-ph/9511376}{hep-ph/9511376}]}.

\bibitem{Mohapatra:1995xd}
R.~N.~Mohapatra and A.~Rasin, ``{Simple supersymmetric solution to the strong
  CP problem},'' \href{https://dx.doi.org/10.1103/PhysRevLett.76.3490}{Phys.\
  Rev.\  Lett.\  {\bfseries 76} (1996) 3490--3493} {\ttfamily
  [\href{https://arxiv.org/abs/hep-ph/9511391}{hep-ph/9511391}]}.

\bibitem{Mohapatra:1996vg}
R.~N.~Mohapatra and A.~Rasin, ``{A Supersymmetric solution to CP problems},''
  \href{https://dx.doi.org/10.1103/PhysRevD.54.5835}{Phys.\  Rev.\  D
  {\bfseries 54} (1996) 5835--5844} {\ttfamily
  [\href{https://arxiv.org/abs/hep-ph/9604445}{hep-ph/9604445}]}.

\bibitem{Balakrishna:1988bn}
B.~S.~Balakrishna and R.~N.~Mohapatra, ``{Radiative Fermion Masses From New
  Physics at Tev Scale},''
  \href{https://dx.doi.org/10.1016/0370-2693(89)91129-5}{Phys.\  Lett.\  B
  {\bfseries 216} (1989) 349--352}.

\bibitem{Babu:1988yq}
K.~S.~Babu and X.~G.~He, ``{DIRAC NEUTRINO MASSES AS TWO LOOP RADIATIVE
  CORRECTIONS},'' \href{https://dx.doi.org/10.1142/S0217732389000095}{Mod.\
  Phys.\  Lett.\  A {\bfseries 4} (1989) 61}.

\bibitem{Ma:2017kgb}
E.~Ma and U.~Sarkar, ``{Radiative Left-Right Dirac Neutrino Mass},''
  \href{https://dx.doi.org/10.1016/j.physletb.2017.08.071}{Phys.\  Lett.\  B
  {\bfseries 776} (2018) 54--57} {\ttfamily
  [\href{https://arxiv.org/abs/1707.07698}{arXiv:1707.07698}]}.

\bibitem{Ma:1989ap}
E.~Ma and G.-G.~Wong, ``{Asymmetric Left-right Model for Generating Radiative
  Quark and Lepton Masses},''
  \href{https://dx.doi.org/10.1103/PhysRevD.41.953}{Phys.\  Rev.\  D {\bfseries
  41} (1990) 953}.

\bibitem{Bonilla:2023wok}
C.~Bonilla, {\em et al.}, ``{Fermion mass hierarchy in an extended left-right
  symmetric model},'' \href{https://dx.doi.org/10.1007/JHEP12(2023)075}{JHEP
  {\bfseries 12} (2023) 075} {\ttfamily
  [\href{https://arxiv.org/abs/2305.11967}{arXiv:2305.11967}]}.

\bibitem{Hall:2023vjb}
L.~J.~Hall, K.~Harigaya, and Y.~Shpilman, ``{Radiative Majorana neutrino masses
  in a parity solution to the strong CP problem},''
  \href{https://dx.doi.org/10.1007/JHEP03(2024)047}{JHEP {\bfseries 03} (2024)
  047} {\ttfamily [\href{https://arxiv.org/abs/2311.10274}{arXiv:2311.10274}]}.

\bibitem{Gabrielli:2016vbb}
E.~Gabrielli, L.~Marzola, and M.~Raidal, ``{Radiative Yukawa Couplings in the
  Simplest Left-Right Symmetric Model},''
  \href{https://dx.doi.org/10.1103/PhysRevD.95.035005}{Phys.\  Rev.\  D
  {\bfseries 95} (2017) 035005} {\ttfamily
  [\href{https://arxiv.org/abs/1611.00009}{arXiv:1611.00009}]}.

\bibitem{deAlmeida:2010qb}
F.~M.~L.~de~Almeida, {\em et al.}, ``{Double seesaw mechanism in a left-right
  symmetric model with TeV neutrinos},''
  \href{https://dx.doi.org/10.1103/PhysRevD.81.053005}{Phys.\  Rev.\  D
  {\bfseries 81} (2010) 053005} {\ttfamily
  [\href{https://arxiv.org/abs/1001.2162}{arXiv:1001.2162}]}.

\bibitem{Brdar:2018sbk}
V.~Brdar and A.~Y.~Smirnov, ``{Low Scale Left-Right Symmetry and Naturally
  Small Neutrino Mass},''
  \href{https://dx.doi.org/10.1007/JHEP02(2019)045}{JHEP {\bfseries 02} (2019)
  045} {\ttfamily [\href{https://arxiv.org/abs/1809.09115}{arXiv:1809.09115}]}.

\bibitem{C:2024exl}
P.~A.~C., {\em et al.}, ``{Left-Right model with radiative double seesaw
  mechanism},'' \href{https://dx.doi.org/10.1007/JHEP12(2024)162}{JHEP
  {\bfseries 12} (2024) 162} {\ttfamily
  [\href{https://arxiv.org/abs/2405.12283}{arXiv:2405.12283}]}.

\bibitem{Nomura:2016run}
T.~Nomura, H.~Okada, and Y.~Orikasa, ``{Radiative neutrino mass in alternative
  left\textendash{}right model},''
  \href{https://dx.doi.org/10.1140/epjc/s10052-017-4657-4}{Eur.\  Phys.\  J.\
  C {\bfseries 77} (2017) 103} {\ttfamily
  [\href{https://arxiv.org/abs/1602.08302}{arXiv:1602.08302}]}.

\bibitem{Babu:2024glr}
K.~S.~Babu and A.~Kaladharan, ``{Dirac Leptogenesis in Left-Right Symmetric
  Models}.'' {\ttfamily
  \href{https://arxiv.org/abs/2410.24125}{arXiv:2410.24125}}.

\bibitem{Bolton:2019bou}
P.~D.~Bolton, F.~F.~Deppisch, C.~Hati, S.~Patra, and U.~Sarkar, ``{Alternative
  formulation of left-right symmetry with $B-L$ conservation and purely Dirac
  neutrinos},'' \href{https://dx.doi.org/10.1103/PhysRevD.100.035013}{Phys.\
  Rev.\  D {\bfseries 100} (2019) 035013} {\ttfamily
  [\href{https://arxiv.org/abs/1902.05802}{arXiv:1902.05802}]}.

\bibitem{Babu:2022ikf}
K.~S.~Babu, X.-G.~He, M.~Su, and A.~Thapa, ``{Naturally light Dirac and
  pseudo-Dirac neutrinos from left-right symmetry},''
  \href{https://dx.doi.org/10.1007/JHEP08(2022)140}{JHEP {\bfseries 08} (2022)
  140} {\ttfamily [\href{https://arxiv.org/abs/2205.09127}{arXiv:2205.09127}]}.

\bibitem{Borah:2017leo}
D.~Borah and A.~Dasgupta, ``{Naturally Light Dirac Neutrino in Left-Right
  Symmetric Model},''
  \href{https://dx.doi.org/10.1088/1475-7516/2017/06/003}{JCAP {\bfseries 06}
  (2017) 003} {\ttfamily
  [\href{https://arxiv.org/abs/1702.02877}{arXiv:1702.02877}]}.

\bibitem{Biswas:2024wbz}
S.~Biswas, V.~P.~K., and A.~Thapa, ``{Connecting pseudo-Nambu-Goldstone dark
  matter with pseudo-Dirac neutrinos in a left-right symmetry model}.''
  {\ttfamily \href{https://arxiv.org/abs/2407.05482}{arXiv:2407.05482}}.

\bibitem{ParticleDataGroup:2024cfk}
{\bfseries Particle Data Group} Collaboration, ``{Review of particle
  physics},'' \href{https://dx.doi.org/10.1103/PhysRevD.110.030001}{Phys.\
  Rev.\  D {\bfseries 110} (2024) 030001}.

\bibitem{Cao:2007rm}
Q.-H.~Cao, E.~Ma, and G.~Rajasekaran, ``{Observing the Dark Scalar Doublet and
  its Impact on the Standard-Model Higgs Boson at Colliders},''
  \href{https://dx.doi.org/10.1103/PhysRevD.76.095011}{Phys.\  Rev.\  D
  {\bfseries 76} (2007) 095011} {\ttfamily
  [\href{https://arxiv.org/abs/0708.2939}{arXiv:0708.2939}]}.

\bibitem{Lundstrom:2008ai}
E.~Lundstrom, M.~Gustafsson, and J.~Edsjo, ``{The Inert Doublet Model and LEP
  II Limits},'' \href{https://dx.doi.org/10.1103/PhysRevD.79.035013}{Phys.\
  Rev.\  D {\bfseries 79} (2009) 035013} {\ttfamily
  [\href{https://arxiv.org/abs/0810.3924}{arXiv:0810.3924}]}.

\bibitem{Okawa:2020jea}
S.~Okawa and Y.~Omura, ``{Light mass window of lepton portal dark matter},''
  \href{https://dx.doi.org/10.1007/JHEP02(2021)231}{JHEP {\bfseries 02} (2021)
  231} {\ttfamily [\href{https://arxiv.org/abs/2011.04788}{arXiv:2011.04788}]}.

\bibitem{Iguro:2022tmr}
S.~Iguro, S.~Okawa, and Y.~Omura, ``{Light lepton portal dark matter meets the
  LHC},'' \href{https://dx.doi.org/10.1007/JHEP03(2023)010}{JHEP {\bfseries 03}
  (2023) 010} {\ttfamily
  [\href{https://arxiv.org/abs/2208.05487}{arXiv:2208.05487}]}.

\bibitem{Planck:2018vyg}
{\bfseries Planck} Collaboration, ``{Planck 2018 results. VI. Cosmological
  parameters},'' \href{https://dx.doi.org/10.1051/0004-6361/201833910}{Astron.\
   Astrophys.\  {\bfseries 641} (2020) A6} {\ttfamily
  [\href{https://arxiv.org/abs/1807.06209}{arXiv:1807.06209}]}. [Erratum:
  Astron.Astrophys. 652, C4 (2021)].

\bibitem{SimonsObservatory:2018koc}
{\bfseries Simons Observatory} Collaboration, ``{The Simons Observatory:
  Science goals and forecasts},''
  \href{https://dx.doi.org/10.1088/1475-7516/2019/02/056}{JCAP {\bfseries 02}
  (2019) 056} {\ttfamily
  [\href{https://arxiv.org/abs/1808.07445}{arXiv:1808.07445}]}.

\bibitem{SimonsObservatory:2019qwx}
{\bfseries Simons Observatory} Collaboration, ``{The Simons Observatory:
  Astro2020 Decadal Project Whitepaper},'' Bull.\  Am.\  Astron.\  Soc.\
  {\bfseries 51} (2019) 147 {\ttfamily
  [\href{https://arxiv.org/abs/1907.08284}{arXiv:1907.08284}]}.

\bibitem{Borah:2025fkd}
D.~Borah, S.~Mahapatra, D.~Nanda, S.~K.~Sahoo, and N.~Sahu, ``{Effective theory
  of light Dirac neutrino portal dark matter with observable ${\Delta N_{\rm
  eff}}$}.'' {\ttfamily
  \href{https://arxiv.org/abs/2502.10318}{arXiv:2502.10318}}.

\bibitem{Bai:2014osa}
Y.~Bai and J.~Berger, ``{Lepton Portal Dark Matter},''
  \href{https://dx.doi.org/10.1007/JHEP08(2014)153}{JHEP {\bfseries 08} (2014)
  153} {\ttfamily [\href{https://arxiv.org/abs/1402.6696}{arXiv:1402.6696}]}.

\bibitem{Chang:2014tea}
S.~Chang, R.~Edezhath, J.~Hutchinson, and M.~Luty, ``{Leptophilic Effective
  WIMPs},'' \href{https://dx.doi.org/10.1103/PhysRevD.90.015011}{Phys.\  Rev.\
  D {\bfseries 90} (2014) 015011} {\ttfamily
  [\href{https://arxiv.org/abs/1402.7358}{arXiv:1402.7358}]}.

\bibitem{Kawamura:2020qxo}
J.~Kawamura, S.~Okawa, and Y.~Omura, ``{Current status and muon $g-2$
  explanation of lepton portal dark matter},''
  \href{https://dx.doi.org/10.1007/JHEP08(2020)042}{JHEP {\bfseries 08} (2020)
  042} {\ttfamily [\href{https://arxiv.org/abs/2002.12534}{arXiv:2002.12534}]}.

\bibitem{Edsjo:1997bg}
J.~Edsjo and P.~Gondolo, ``{Neutralino relic density including
  coannihilations},'' \href{https://dx.doi.org/10.1103/PhysRevD.56.1879}{Phys.\
   Rev.\  D {\bfseries 56} (1997) 1879--1894} {\ttfamily
  [\href{https://arxiv.org/abs/hep-ph/9704361}{hep-ph/9704361}]}.

\bibitem{Saikawa:2020swg}
K.~Saikawa and S.~Shirai, ``{Precise WIMP Dark Matter Abundance and Standard
  Model Thermodynamics},''
  \href{https://dx.doi.org/10.1088/1475-7516/2020/08/011}{JCAP {\bfseries 08}
  (2020) 011} {\ttfamily
  [\href{https://arxiv.org/abs/2005.03544}{arXiv:2005.03544}]}.

\bibitem{Belanger:2020gnr}
G.~Belanger, A.~Mjallal, and A.~Pukhov, ``{Recasting direct detection limits
  within micrOMEGAs and implication for non-standard Dark Matter scenarios},''
  \href{https://dx.doi.org/10.1140/epjc/s10052-021-09012-z}{Eur.\  Phys.\  J.\
  C {\bfseries 81} (2021) 239} {\ttfamily
  [\href{https://arxiv.org/abs/2003.08621}{arXiv:2003.08621}]}.

\bibitem{Abe:2015rja}
T.~Abe and R.~Sato, ``{Quantum corrections to the spin-independent cross
  section of the inert doublet dark matter},''
  \href{https://dx.doi.org/10.1007/JHEP03(2015)109}{JHEP {\bfseries 03} (2015)
  109} {\ttfamily [\href{https://arxiv.org/abs/1501.04161}{arXiv:1501.04161}]}.

\bibitem{Higuchi:2023kbt}
R.~Higuchi, S.~Iguro, S.~Okawa, and Y.~Omura, ``{Light mass window of inert
  doublet dark matter with lepton portal interaction},''
  \href{https://dx.doi.org/10.1103/PhysRevD.109.075007}{Phys.\  Rev.\  D
  {\bfseries 109} (2024) 075007} {\ttfamily
  [\href{https://arxiv.org/abs/2310.13685}{arXiv:2310.13685}]}.

\bibitem{LZ:2022lsv}
{\bfseries LZ} Collaboration, ``{First Dark Matter Search Results from the
  LUX-ZEPLIN (LZ) Experiment},''
  \href{https://dx.doi.org/10.1103/PhysRevLett.131.041002}{Phys.\  Rev.\
  Lett.\  {\bfseries 131} (2023) 041002} {\ttfamily
  [\href{https://arxiv.org/abs/2207.03764}{arXiv:2207.03764}]}.

\bibitem{PandaX-4T:2021bab}
{\bfseries PandaX-4T} Collaboration, ``{Dark Matter Search Results from the
  PandaX-4T Commissioning Run},''
  \href{https://dx.doi.org/10.1103/PhysRevLett.127.261802}{Phys.\  Rev.\
  Lett.\  {\bfseries 127} (2021) 261802} {\ttfamily
  [\href{https://arxiv.org/abs/2107.13438}{arXiv:2107.13438}]}.

\bibitem{DEAP-3600:2017ker}
{\bfseries DEAP-3600} Collaboration, ``{Design and Construction of the
  DEAP-3600 Dark Matter Detector},''
  \href{https://dx.doi.org/10.1016/j.astropartphys.2018.09.006}{Astropart.\
  Phys.\  {\bfseries 108} (2019) 1--23} {\ttfamily
  [\href{https://arxiv.org/abs/1712.01982}{arXiv:1712.01982}]}.

\bibitem{DEAP-3600:2017uua}
{\bfseries DEAP-3600} Collaboration, ``{First results from the DEAP-3600 dark
  matter search with argon at SNOLAB},''
  \href{https://dx.doi.org/10.1103/PhysRevLett.121.071801}{Phys.\  Rev.\
  Lett.\  {\bfseries 121} (2018) 071801} {\ttfamily
  [\href{https://arxiv.org/abs/1707.08042}{arXiv:1707.08042}]}.

\bibitem{DarkSide-50:2022qzh}
{\bfseries DarkSide-50} Collaboration, ``{Search for low-mass dark matter WIMPs
  with 12~ton-day exposure of DarkSide-50},''
  \href{https://dx.doi.org/10.1103/PhysRevD.107.063001}{Phys.\  Rev.\  D
  {\bfseries 107} (2023) 063001} {\ttfamily
  [\href{https://arxiv.org/abs/2207.11966}{arXiv:2207.11966}]}.

\bibitem{DarkSide-20k:2017zyg}
{\bfseries DarkSide-20k} Collaboration, ``{DarkSide-20k: A 20 tonne two-phase
  LAr TPC for direct dark matter detection at LNGS},''
  \href{https://dx.doi.org/10.1140/epjp/i2018-11973-4}{Eur.\  Phys.\  J.\  Plus
  {\bfseries 133} (2018) 131} {\ttfamily
  [\href{https://arxiv.org/abs/1707.08145}{arXiv:1707.08145}]}.

\end{thebibliography}\endgroup

\end{document}